\documentclass[preprint,12pt]{elsarticle}

\usepackage{amssymb}
\usepackage{graphicx}
\usepackage{svg}
\graphicspath{{figs}}

\usepackage{ulem}
\usepackage{amsmath}
\usepackage{amssymb}
\usepackage[utf8]{inputenc}
\usepackage{amsmath}
\usepackage{longtable}
\usepackage{multirow}
\usepackage{multicol}

\usepackage[linesnumbered,boxed]{algorithm2e}
\SetKwInput{KwInput}{Input}
\SetKwInput{KwOutput}{Output}

\usepackage{url}
\usepackage{subfiles} %

\journal{Computers in Biology and Medicine}

\begin{document}

\begin{frontmatter}

\title{MuLaN: a MultiLayer Networks Alignment Algorithm}

\author[1,4,cor]{Marianna Milano\corref{contrib}\corref{corauthor}}
\author[2,4]{Pietro Cinaglia\corref{contrib}}
\author[3,4]{Pietro Hiram Guzzi\corref{corauthor}}
\author[3,4]{Mario Cannataro}

\cortext[corauthor]{Corresponding author.} 
\cortext[contrib]{Authors contributed equally.}

\address[1]{Department of Experimental and Clinical Medicine, Magna Graecia University of Catanzaro, 88100 Catanzaro, Italy.}

\address[2]{Department of Health Sciences, Magna Graecia University of Catanzaro, 88100 Catanzaro, Italy.}

\address[3]{Department of Medical and Surgical Sciences, Magna Graecia University of Catanzaro, 88100 Catanzaro, Italy.}

\address[4]{Data Analytics Research Center, Magna Graecia University of Catanzaro, 88100 Catanzaro, Italy.}

\begin{abstract}
A Multilayer Network (MN) is a system consisting of several topological levels (i.e., layers) representing the interactions between the system's objects and the related interdependency. Therefore, it may be represented as a set of layers that can be assimilated to a set of networks of its own objects, by means \textit{inter}-layer edges (or \textit{inter}-edges) linking the nodes of different layers; for instance, a biological MN may allow modeling of \textit{inter} and \textit{intra} interactions among diseases, genes, and drugs, only using its own structure.
The analysis of MNs may reveal hidden knowledge, as demonstrated by several algorithms for the analysis. Recently, there is a growing interest in comparing two MNs by revealing local regions of similarity, as a counterpart of Network Alignment algorithms (NA) for simple networks. However, classical algorithms for NA such as Local NA (LNA) cannot be applied on multilayer networks, since they are not able to deal with \textit{inter}-layer edges. Therefore, there is the need for the introduction of novel algorithms.
In this paper, we present \textit{MuLaN}, an algorithm for the local alignment of multilayer networks. We first show as proof of concept the performances of \textit{MuLaN} on a set of synthetic multilayer networks. Then,  we used as a case study a real multilayer network in the biomedical domain.
Our results show that \textit{MuLaN} is able to build high-quality alignments and can extract knowledge about the aligned multilayer networks. 
\\
\textit{MuLaN} is available at \url{https://github.com/pietrocinaglia/mulan}.

\end{abstract}




\end{frontmatter}


\section{Introduction}
\label{sec:sample1}
\begin{figure}[!ht]
\centering
\includegraphics[width=0.8\textwidth]{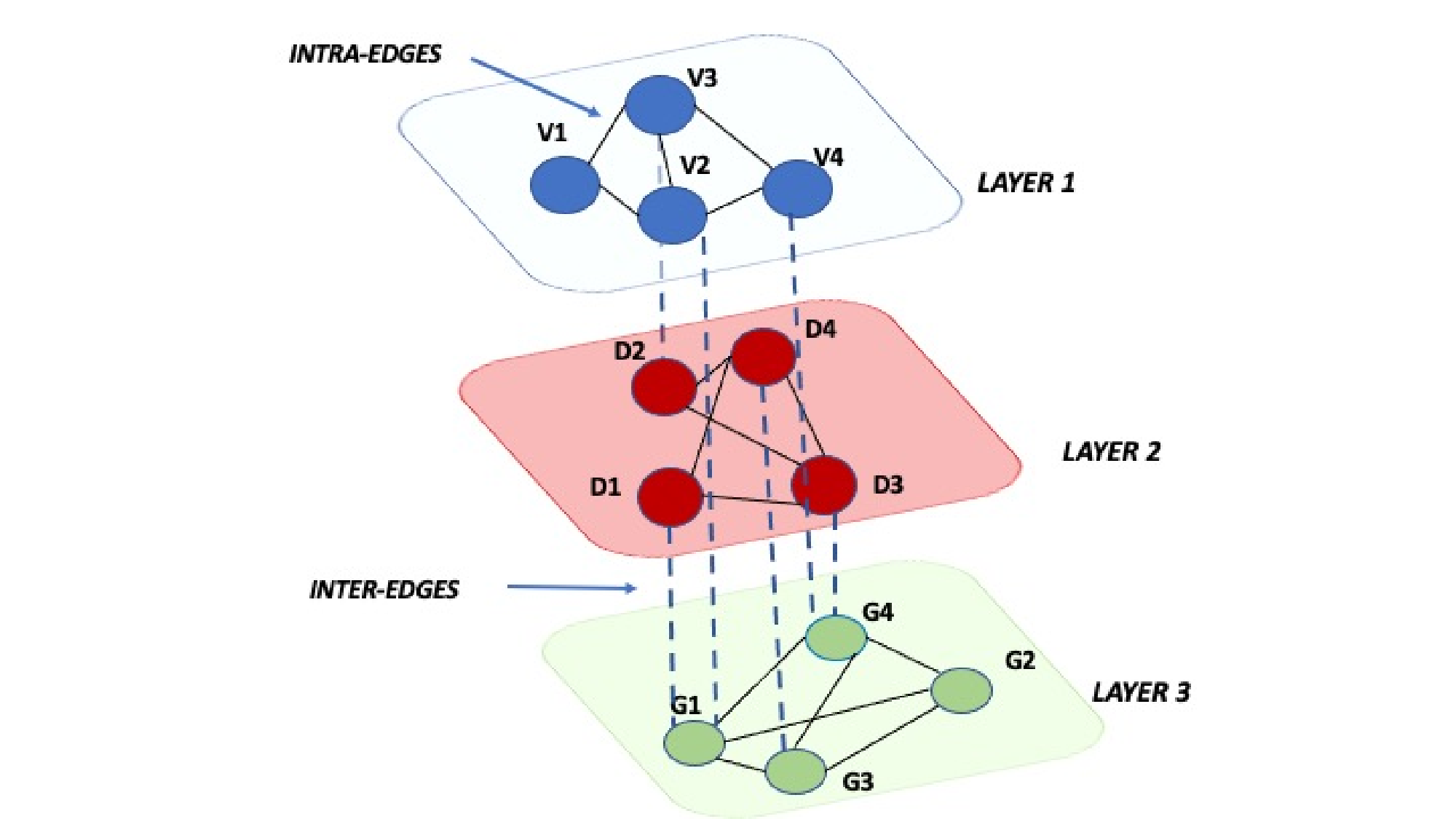}
\caption{A non-exhaustive example of multilayer network. The figure represents a network in a multilayer prospective. The same entities are modeled as nodes of a same layer, and their interactions consists of \textit{intra}-edges. Otherwise, the nodes of different layers are connected by \textit{inter}-edges.}
\label{fig:diap1}
\end{figure}

\begin{figure}[!ht]
\centering
\includegraphics[width=0.8\textwidth]{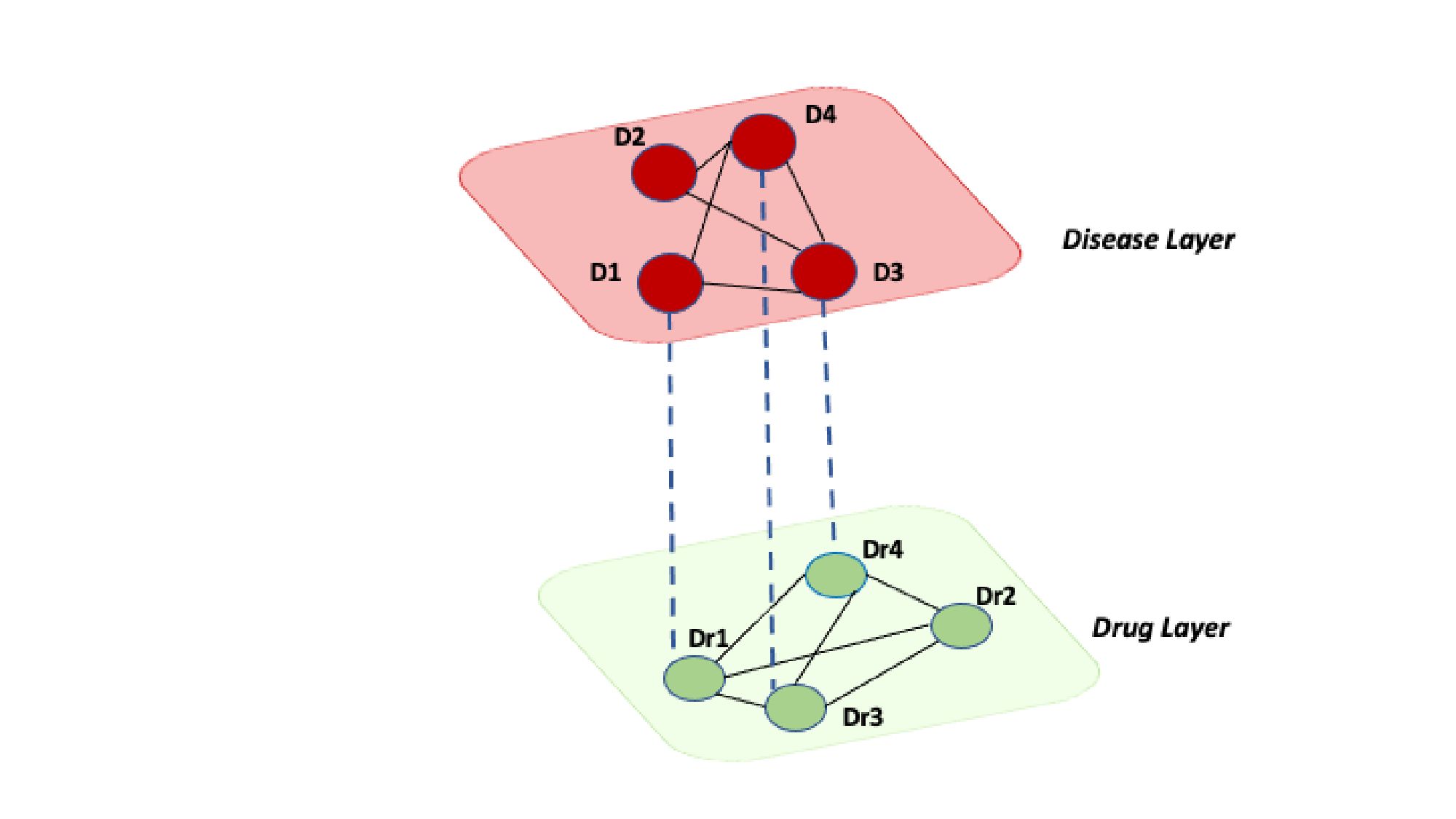}
\caption{The figure shows a toy example of a biological multilayer network, representing disease-drug associations. The nodes are the diseases and the drugs, both discriminated by belonging to the respective layer. The \textit{intra}-edges represent the drug-drug and the disease-disease associations, while the \textit{inter}-edges the disease-drug associations. For instance, these may be analyzed to investigate the mechanisms underlying the interaction between diseases and drugs, or for computational drug repositoring.}
\label{fig:diap12}
\end{figure}

Networks are largely used to represent entities and their association in many fields \cite{cannataro2010protein,barabasi2004network,gu2022modeling}. For instance, in computational biology and bioinformatics, networks are used to model the set of association among genes, proteins and other macromolecules \cite{milano2017extensive,dondi2021novel,eskandarzade2022network}. More recently, it has been shown that a single level of representation, i.e., considering only genes, or only proteins, may result in a loss of information, so the need for introduction of other models arise. Such models are based on a multilevel (or multilayered) view of a system. As an example, in the biological scenario, each level may correspond to a different view, e.g. genes, proteins, and diseases. In such a view, each layer is composed by a set of homogeneous nodes linked by edges (i.e., \textit{intra}-layer edges). Associations between two different layers are modelled by cross layer edges (i.e., \textit{inter}-layer edges) \cite{tagarelli2017ensemble,hammoud2020multilayer}.
Figure \ref{fig:diap1} represents a non-exhaustive example of a simple multilayer network having three layers. Each layer is a different network. Figure \ref{fig:diap12} depicts an example of a biological multilayer network representing the interplay between diseases and drugs.

Formally, a multilayer graph may be described as a tuple $G_{M} = (V_M, E_M, V, L)$, where $V_M$ and $E_M$ are a set of nodes and edges, respectively, and $L$ is a set of layers. 
Thus, given $L=\{L_1,L_2,...,L_k\}$, with $k$ the number of layers in the multilayer graph, the nodes as pairs $V_M \subseteq V x L_1 x ... L_k$, the edges $E_M \subseteq V_M x V_M $ connect the pairs $(v, l)$, $(v', l')$.
In a multilayer graph, an edge is defined \textit{intra}-layer in case of $l$ is equal to $l'$ or \textit{inter}-layer, in case of $l$ and $l'$ are different \cite{kivela2014multilayer}.





In general, the analysis of networks is rarely simple.
Common challenges that can be generalized for all types of networks concern i) the modelling of the network, i.e., what it represents and ii) the analysis of the network, i.e., the metrics grabbing  the biological phenomenon of interest. 
These aspects are at least as important for multilayer networks, given their added complexity.
Furthermore, there are additional observations that are different to multilayer networks with respect to classical networks. For instance, an analysis should take into account the difference among layers. 
In particular, since the strength of multilayer network analysis consists of  the capability to include information on different type of relationship, it would be necessary to consider which layers should be included in the analysis, and the interpretation of  intra-layer and inter-layer edges values, since they discriminate against the different relationships. 

The  analysis of networks enables to find interesting regions of similarity between them \cite{cinagliaCannataroSurvey,agapito2013coresnp}. For simple networks, this task is accomplished by means of Network Alignment (NA) algorithms. Unfortunately, as also evidenced in \cite{milano2020hetnetaligner,kumar2021data,milano2022design}, existing alignment algorithms are unable to process multilayer networks. Therefore, the need for defining a novel multilayer network algorithm arises. We focused in particular on evidencing relatively small regions of similarity among the multilayer networks, so we extended the Local Network Algorithm approaches. In a previous work \cite{milano2022design}, we presented \textit{MultiLoAl} for the local alignment of multilayer networks. Despite the effectiveness of the approach, it presented some limitations related to the scalability on large datasets and on the topological structure of the region of the similarity. 

Therefore, we extended our methodology to admit further topological structures of the found regions as well as better to obtain performances in terms of quality and running time. Such a methodology has been implemented in a novel tool, namely \textit{MuLaN} (from \textit{MultiLAyer Network Alignment Algorithm}). It consists of a novel approach for the building of alignment graph \footnote{the alignment graph is the key data structure for local alignment algorithm}  who achieves better performances, and it is integrated with a set of community discover algorithms enabling to discover regions of similarity with different topological structures.


Summarizing, \textit{MuLaN} allows you to mainly benefit from the following advantages, compared to \textit{MultiLoAl}: (i) a reduction of running time for building the alignment graph, and (ii) a deeper experimentation on a larger dataset.

\begin{figure*}[!ht]
 \centering
 \includegraphics[width=0.75\textwidth]{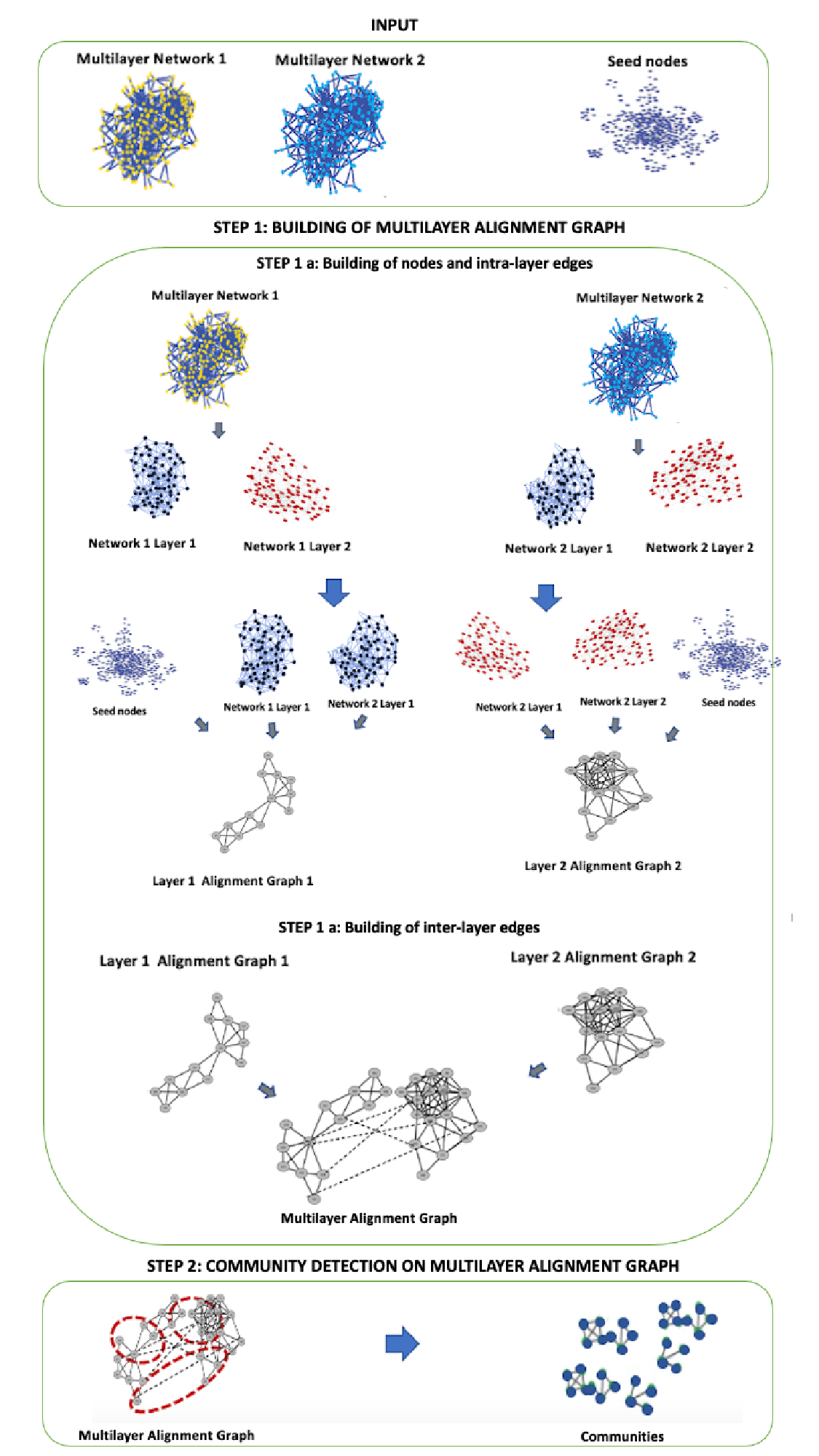}
 \caption{Workflow of the proposed algorithm.}
 \label{fig:workflow}
\end{figure*}

\textit{MuLaN} is based on the workflow depicted in Figure \ref{fig:workflow}. 

It receives as input two multilayer networks ($M_1$ and $M_2$) having the same number of layers $n$. Without loss of generality, we suppose the existence of a bijective correspondence among layers of the networks, so the layer $k$ in 1-$n$ of $M_1$ corresponds to the layer $k$ of $M_2$. \textit{MuLaN} also receives a multiset of seed nodes representing the similarity between the node of the graph of the same layer $k$ of the two networks. Starting from $M_1$ and $M_2$ and multiset of seed nodes, at first, \textit{MuLaN} analyzes each pair of $k$ layers in  $M_1$  and $M_2$ independently. Then, it builds $n$ alignment graphs. After building each alignment graph for each $k$ layer separately, \textit{MuLaN} analyzes $M_1$ and $M_2$ to add \textit{inter}-layer edges among $n$ alignment graphs. Thus, the algorithms build $n-1$ alignment graphs, that we call  multilayer alignment graphs. Finally, \textit{MuLaN} applies a community detection algorithm suitable for multilayer networks on the final multilayer alignment graph to  detect communities representing local regions of similarity, i.e. a single region of the local alignment. The current version of \textit{MuLaN} offer three different community discover algorithms: \textit{Louvain} \cite{jutla2011generalized},  \textit{Infomap} \cite{de2015identifying} and \textit{Greedy} \cite{newman2004fast}.

We implemented three versions of \textit{MuLaN} by varying the community detection method. The default version applies \textit{Louvain} \cite{jutla2011generalized} to mine community from the multilayer alignment graph, whereas, the other two versions use \textit{Infomap} \cite{de2015identifying} and \textit{Greedy} \cite{newman2004fast} as community detection algorithms, respectively \cite{magnani2021community}.

The rest of this paper is organized as follows. Section \ref{section:relatedworks} discusses the background on multilayer networks and multilayer community detection, Section \ref{section:methods} presents the \textit{MuLaN} Algorithm, and Section \ref{section:resultsDiscussion} presents and discusses the results. Finally, Section \ref{section:conclusions} concludes the paper.
\section{Related Work}
\label{section:relatedworks}

We report here some existing state-of-the-art approaches for the analysis of MNs. We separate the algorithms that analyze a single network from those comparing two networks \cite{guzzi2018survey}.

\subsection{Network Alignment Algorithms}

The problem of graph alignment consists of finding a mapping between nodes of two or more graphs to maximize an associated cost function. Formally, let $G_{1}= \{V_{1},E_{1}\}$ and $G_{2}= \{V_{2},E_{2}\}$ two graphs, where $V_{1,2}$ are sets of nodes and $E_{1,2}$ are sets of edges, the \textbf{graph alignment problem} consists of finding a function (or a mapping) $f: V_1\rightarrow V_2$ such that the similarity between mapped entities is maximized. Among the others, algorithms may be categorised in global network alignment (GNA) algorithms (i.e. algorithms that aim to find a single mapping among all the nodes of the networks), or local network alignment (LNA) algorithms (i.e. algorithms aiming at finding multiple small regions of similarity among networks).

We here report some LNA algorithms, since MuLaN search for multiple regions of similarity in multilayer networks. 
 
An example of LNA algorithm is AlignNemo \cite{ciriello2012alignnemo} that enables the discovery of subnetworks of proteins related to biological function and topology. 
AlignMCL \cite{mina2014improving} extends AlignNemo \cite{ciriello2012alignnemo}, by providing a formal definition of  the alignment graph, which is clustered by using the Markov cluster algorithm MCL \cite{enright2002efficient}.

Also, GLAlign (Global Local Aligner) \cite{milano2018glalign} combines the topology information from global alignment with biological information (e.g. homology relationships) to build local network alignment. 
NetworkBLAST \cite{SharanIdeker2006} aims to find small dense regions in protein-protein interaction networks. Such subgraphs represent protein complexes, i.e. groups of proteins performing a similar function or involved in the same biological process.  
NetAligner \cite{pache2012netaligner} presents  a strategy to identify evolutionarily conserved interactions on the basis of the consideration that interacting proteins evolve at rates significantly closer than expected by chance.
Furthermore, L-HetNetAligner \cite{milano2020hetnetaligner} extends the local alignment to the heterogeneous networks. Finally, some recent papers focused on the alignment of dynamic or dual networks \cite{dante,guzzi2020extracting,dondi2021novel}. However, these network alignment algorithms do not perform very well for multilayer networks \cite{tagarelli2017ensemble,ren2021pattern}.

\subsection{Community Detection in Multilayer Networks}
Community detection algorithms aim to identify groups of nodes closely connected with respect to the average of the network, starting from the assumption that nodes in the same community have a similar  role \cite{fortunato2016community, lancichinetti2012consensus, gligorijevic2016fusion}.
In detail, a community is defined as groups of nodes that are more densely connected than the rest of the network; they represent significant characteristics for understanding the functionalities and organizations of complex systems modeled as a network. 

In multilayer networks, the communities represent groups of well-connected nodes considering whole layers, community detection methods should consider the diversity between the layers\cite{magnani2021community}.


To handle these issues, many community detection algorithms for multilayer networks have been recently developed.

The \textit{Louvain} algorithm, initially developed for simple networks, has been extended to handle multilayer networks \cite{jutla2011generalized}. The algorithms are based on an ad hoc defined notion of modularity for multilayer network, which is based on the definition of community as a set of nodes and edges belonging to multiple layers. Therefore, the algorithm applies an iterative greedy approach to build communities.




\textit{Infomap} \cite{de2015identifying,rosvall2008maps} discovers communities in multilayer networks by running random walks, admitting also interlayer edges. \textit{Infomap} can be used to find both overlapping and non-overlapping communities.


\textit{Greedy} \cite{newman2004fast} algorithm is a simple and fast community detection method that works by greedily optimizing a quality function known as modularity.
The algorithm starts with each node in its own community and iteratively merges communities that increase the modularity at the most.

ABACUS \cite{berlingerio2013abacus} algorithm enables to extract multidimensional communities by mining frequent closed item sets on one-dimensional community memberships.
Initially, it extracts one-dimensional communities, looking at each dimension. Then, ABACUS assigns a label to nodes that consists in a list of pair tags (e.g., dimension and community).
After that, ABACUS uses a frequent closed item set mining algorithm by treating each pair of tags as an item. At the end, the multidimensional communities described by the item sets consist of frequent closed item sets.

Multi-Layer Clique Percolation \cite{afsarmanesh2018partial} method applies the classical clique percolation method on traditional networks. In the classical version, dense regions are cliques, whereas the two cliques are adjacent if they have common nodes. Multi-Layer Clique Percolation extends research of cliques by considering multiple layers, and it redefines the adjacency metric by considering both common nodes and common layers are expected. Thus, the communities correspond to the combinations of adjacent cliques.

Multi-Dimensional Label Propagation  \cite{boutemine2017mining} is an extension of the Label Propagation algorithm, which is a semi-supervised learning algorithm used for graph-based classification problems. The MDLP algorithm is designed to work with multilayer networks.

In MDLP, each layer of the multilayer network is treated as a separate graph, and the algorithm propagates labels across all layers simultaneously. It allows incorporating information from multiple sources and to capture complex relationships between nodes across layers. Furthermore, it can be computationally expensive, especially for large multilayer networks, and may require careful parameter tuning to achieve optimal results.
\section{\textit{MuLaN} Algorithm}
\label{section:methods}

The \textit{MuLaN} algorithm builds the alignment through two steps:

\begin{itemize}
 \item \textbf{Building of the Multilayer Alignment Graph}: 
 First, it integrates the input networks into a multilayer  alignment graph (MAG) using a set of initial pairs of correspondences;
 \item \textbf{Analysis of the Multilayer Alignment Graph}: Second, it reveals the regions of similarity by analysing the topology of the multilayer alignment graph. 
\end{itemize}

The core of the algorithm is the multilayer alignment graph, that is used to integrate initial input networks and which is able to contains the regions of similarity. The multilayer alignment graph is a multilayer network which has the same layer of the input networks. Each layer ${K}$ of the MAG is an alignment graph between the corresponding pairs of layers of the input network. The edges between two layers of MAG are derived by analysing inter-layer edges of the input networks, as shown in the following.





Figure \ref{fig:workflow} shows the workflow of the algorithm, while Algorithm 1 shows the pseudocode of \textit{MuLaN}.

\begin{algorithm}
\KwInput{$G_1=(V1,E_1, C_1)$, and $G_2=(V2,E_2, C2)$,$\Delta$, A set of high-similar seed nodes}
\KwResult{A set of Aligned Regions}
\textit{Initialization}\;
\textbf{Step 1 -- Building of the Multilayer Alignment Graph }: $G{al1}_=(V_{al},E_{al})$ \\ $G{al2}_= (V_{al}, E_{al})$\\
\ForAll{Pair of the input list of paired nodes}{add a node in Alignment Graph}
\ForAll{Nodes in $V_{al}$}{Add Edges Verifying Match-Mismatch- Conditions}
Building of the \textit{inter}-layer Alignment Graph : $G{inter_l}=(G{1_al},G{2_al},E_{inter_l})$\\
\textbf{Step 2 -- Community Detection Algorithm on:} $G{inter_l}=(G{1_al},G{2_al},E_{inter_l})$.\\
\Return{A set of subgraphs of $G{inter_l}=(G{1_al},G{2_al},E_{inter_l})$
}

\caption{\textit{MuLaN} Algorithm}
\end{algorithm}

\begin{figure}[!ht]
\centering
\includegraphics[width=0.8\textwidth]{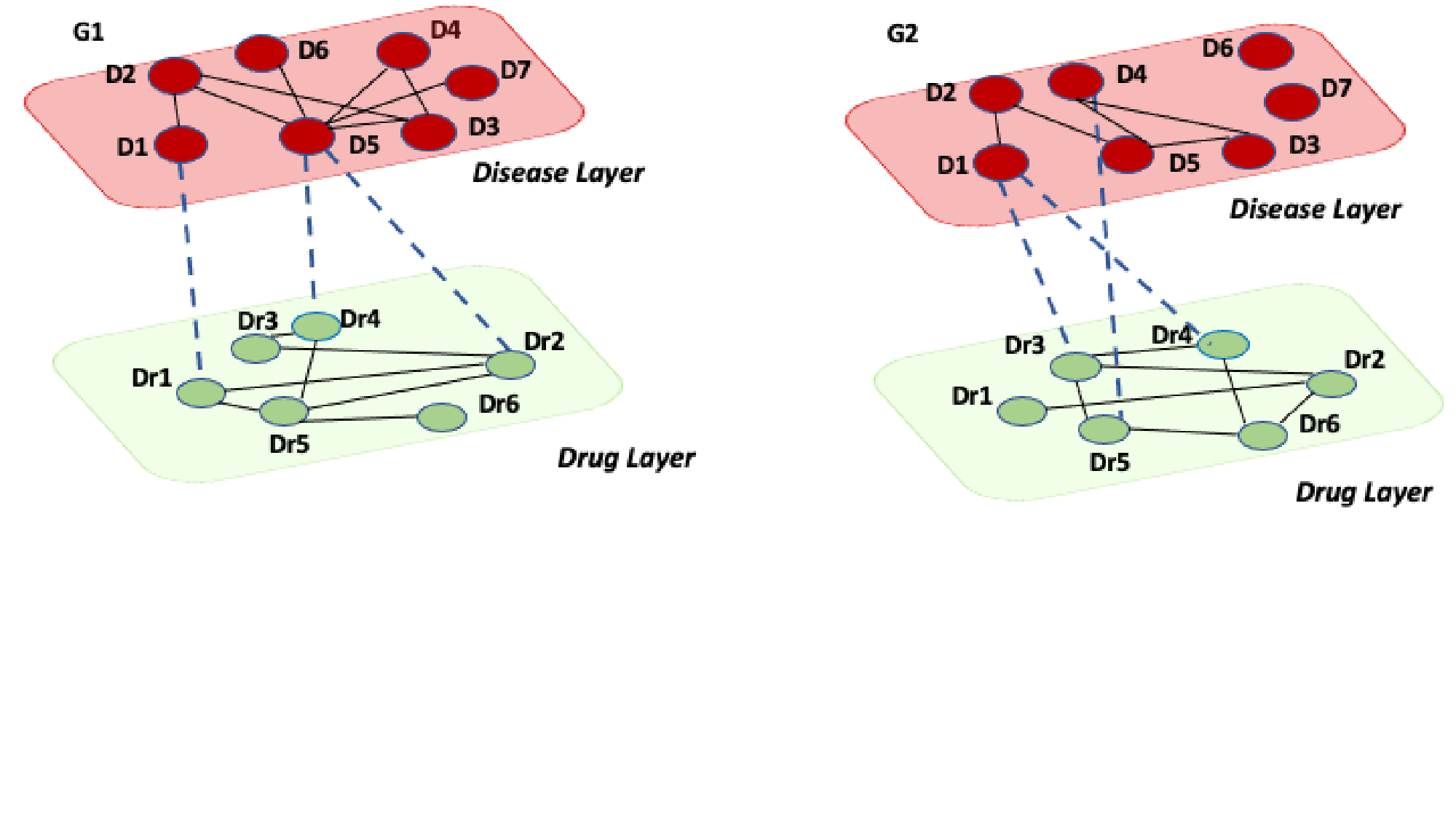}
\caption{Example of two Multilayer networks. The nodes represent a set of diseases and a set of drugs that belong to the Drug layer and the Disease layer. Please note that each layer networks has different edges; moreover, also, the inter-layer edges are different.}
\label{fig:diap6}
\end{figure}

We explain our contribution using a toy example of a multilayer network with two layers without loss of generality; then we will discuss its formalization.
Let us consider two multilayer input networks $G_1$, and $G_2$ presenting two layers: disease and drug, as reported in Figure \ref{fig:diap6}.
Node colors are used to distinguish different types of nodes belonging to two different types of layers. 
For simplicity, the two multilayer input networks have the same number of nodes.

\subsection{ Step 1: Building of the multilayer alignment graph}

\subsection{Step (1.a). Building of \textit{intra}-layer edges.}
In the first step, MuLaN builds the MAG in two sub steps, initially it build all the alignment graphs for each layer $k$ (step 1.a, see Appendix for complete details), then it add all the \textit{inter}-layer edges (Step 1.b).

\subsection{Step (1.b). Building of \textit{inter}-layer edges.}
In the step 1.b, the algorithm adds the 
\textit{inter}-layer edges. This step considers all the pairs of layers in the MAG. For each pair of layers ${k,t}$, it analyzes all the nodes. Each node of the graph into a single layer represents a pair of nodes in the corresponding of nodes. 
For each pair of nodes of the layers, \textit{MuLaN} analyzes the corresponding layer of the input networks, and it inserts and weights the edges considering two conditions: \textbf{match}, or \textbf{mismatch}.

\begin{figure}[!ht]
\centering
\includegraphics[width=0.8\textwidth]{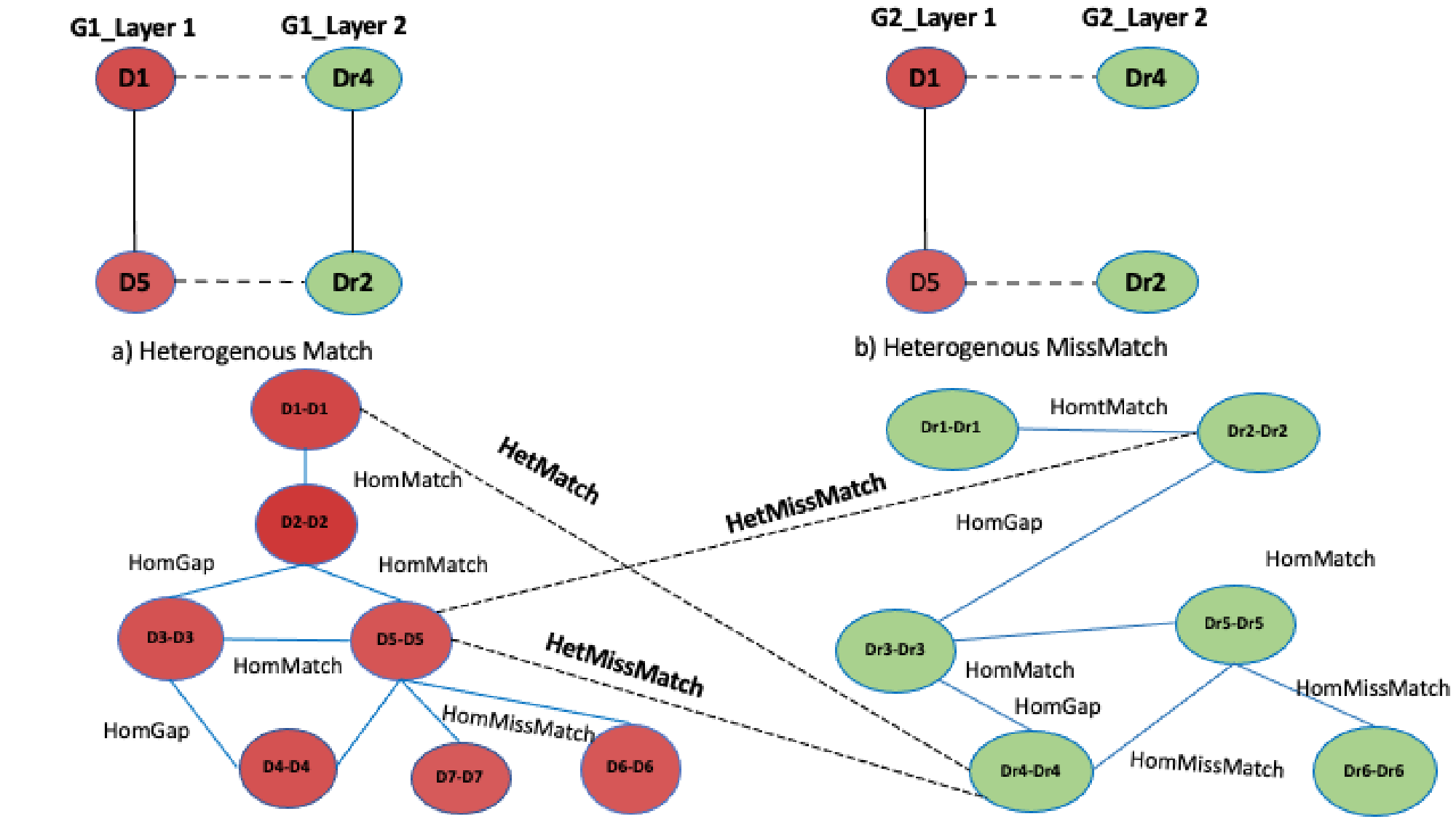}
\caption{Example of a heterogeneous match, mismatch and building of \textit{inter}-layer Alignment Graph.}
\label{fig:diap8}
\end{figure}

Let us consider the nodes of the alignment graphs 1 and the alignment graphs 2; in particular, let us analyze the pair of nodes $(D1-Dr4)$ and $(D5-Dr2)$ in Figure \ref{fig:diap6}. To determine the presence of an edge, we consider the edges $(D1,Dr4) \in G_1$ network and $(D1,Dr4) \in G_2$ network.
If $G_1$, and $G_2$ contains these nodes, and the nodes are adjacent, there is a \textbf{match}, that we call \textbf{heterogeneous match}
because the node type are different, see Figure \ref{fig:diap8} (a).

Otherwise, if $G_1$, and $G_2$ contain these nodes, and the nodes are adjacent only in a single network, there is a \textbf{mismatch} which we call \textbf{heterogeneous mismatch} Figure \ref{fig:diap8} (b). 

Finally, the algorithm assigns the weight to each edge as follows: heterogeneous match equal to 0.9, heterogeneous mismatch equal to 0.4.
At the end of this step, the multilayer alignment graph is built.

\subsection{Step 2. Community Detection on the multilayer alignment graph}

At this point, the algorithms analyzes the MAG to discover communities employing one of the previously introduced community detection methods \cite{kim2015community, paul2020spectral, guzzi2020extracting, huang2021survey}. Since our methodology presents a general design,  it is possible to mine the MAG by applying  different community detection methods. 

In the current version of \textit{MuLaN}, we applied \textit{Louvain} algorithm to mine the communities on the alignment graph. However, the user can choose the community detection algorithm by selecting among \textit{Louvain}, \textit{Infomap} and \textit{Greedy} algorithms.

\textit{MuLaN} generates three outputs: 1) a file containing the mined communities; 2) a file of the multilayer alignment graph represented as edge list, where the weight of the edge according to homogeneous/heterogeneous match, homogeneous/mismatch, homogeneous gap cases and type of layer; 3) a file containing the number of mined communities, the modularity value (where modularity is the measure related to the network structure that detects the density of connections within a module) and the run time (see an example of output at \url{https://github.com/pietrocinaglia/mulan}). 

\begin{figure}[!ht]
\centering
\includegraphics[width=0.8\textwidth]{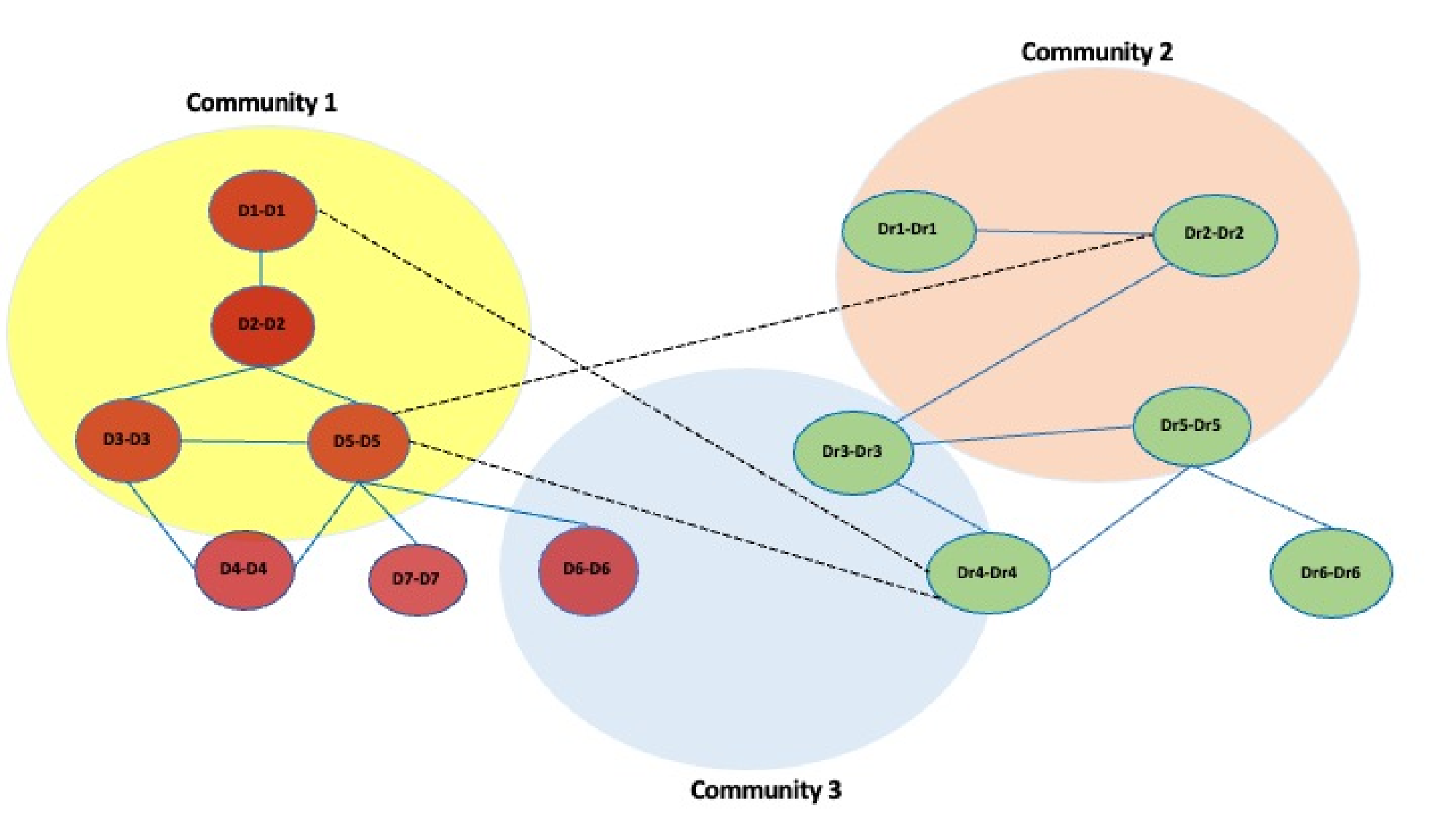}
\caption{Example of Community Detection Extraction on \textit{inter}-layer Alignment Graph.} 
\label{fig:diap5}
\end{figure}
\subsection{Complexity of the algorithm}
The asymptotic temporal complexity of the alignment process was estimated as $NxM$, with $N$ the number of interactions defined into the source network and $M$ that of the target one, calculated on the basis of the respective edge list.

The latter was designed as a multi-line adjacency list with data values, in order to store both node and edge attributes: node type, and the layer upon which (or between) the interaction is modeled. This approach allows us to have a flat view of the graph, which is therefore aligned using the same basic principles of alignment between static networks; its complexity will be proportionate to the resulting size.

It turns out to be quadratic, for multilayer networks consisting of edge lists with a same size ($N=M$).
\section{Results and Discussion}
\label{section:resultsDiscussion}


As a proof of principle, we experimented \textit{MuLaN} on two datasets, consisting of (i) ten synthetic multilayer networks, and (ii) one  real multilayer network.

\subsection{Dataset 1: Synthetic Networks}

We build ten multilayer networks with two layers. Each layer is a Barabási-Albert network \cite{barabasi2004} having $1000$ nodes ($n=1000$) and $1000$ edges ($m=1$, with $m$ the number of edges to attach from a new node to existing ones).

The number of \textit{inter}-edges was defined as $30\%$ (i.e., $300$) of the \textit{intra}-edges, and these were randomly generated. The resulting edge list consisted of $2300$ interactions among $2000$ nodes, for the whole network.


According to the described approach, we modeled a set of $10$ initial multi-layer networks with $2$ layers, and five noisy versions for each one.  The noisy version was built by removing $5\%$, $10\%$, $15\%$, $20\%$, and $25\%$ of interactions, randomly selected from the set of \textit{inter}- and \textit{intra}- edges. We have generated the pairs to be aligned by using the initial networks and the respective noisy versions.


\subsection{Dataset 2: Real Network}

We considered the following datasets of the Stanford Biomedical Network Dataset Collection (BioSNAP) \cite{biosnapnets}:


\begin{itemize}
 \item \textit{Drug-Drug Interaction (DDI)} network of interactions between drugs, approved by the U.S. Food and Drug Administration (FDA): $1514$ nodes and $48514$ edges.
 \item \textit{Disease-Disease (DD)} network of interaction between inherited diseases: $6878$ nodes and $6877$ edges.
 \item \textit{Disease-Drug Association (DDA)} network, a set of curated relationships between diseases and drugs: $5535$ disease nodes, $1662$ chemical nodes, and $466656$ edges.
\end{itemize}

We build a multilayer network with two layers obtained from the DDI and DD database. Then, we added inter-layer edges by considering the DDA database. Subsequently, a preprocessing was also performed to remove (i) zero degree nodes, (ii) duplicate edges, and (iii) objects outside the intersection set from the DDI, DD, and DDA networks. Finally, the resulting multilayer network consisted of $83,92$ nodes and $128,200$ interaction, of which $72,809$ \textit{inter}-edges.

Similarly to our own synthetic networks, we generated five noised versions of the real network by removing $5\%$, $10\%$, $15\%$, $20\%$, and $25\%$ of interactions, randomly selected from the set of \textit{inter}- and \textit{intra}- edges.

\subsection{Alignment Parameters}
We set $\Delta $distance equal to 2, and following weights:
\begin{itemize}
	\item Homogeneous Match : 1
	\item Homogeneous Mismatch: 0.5
          \item Homogeneous Gap: 0.2
         \item Heterogeneous Match: 0.9
	\item Heterogeneous Mismatch: 0.4.
 \end{itemize}
\subsection{Performance Evaluation}
We evaluated \textit{\textit{MuLaN}} alignments by considering:

\begin{itemize}
 \item the alignment of a network with respect to itself to show the ability to find known regions of similarity;
 \item the alignment of the network with respect to an altered version of the network obtained by adding different levels of noise ($5\%$, $10\%$, $15\%$, $20\%$, and $25\%$) by randomly removing edges from the network.
\end{itemize}

We aimed to demonstrate the ability of our algorithm to build high-quality alignments with edge conservation of about $90\%$.

The proposed solution aligned high-confidence synthetic networks to themselves and to their noisy counterparts. Overall, we computed a set of $60$ local alignments.

In addition to the experimentation on synthetic data, we aligned the high-confidence real network with itself and its noisy counterparts, by producing a total of $6$ local alignments.

All experiments have been conducted on a 64-bit workstation with the following specifications: AMD Ryzen 3 (2.6 GHz Dual-Core) and 8 GB of RAM.

\textit{MuLaN} aligned two multilayer networks consisting of $2000$ nodes and $2298$ edges, each one, in $\sim1.25$ seconds. To the latter, it was then necessary to include the time required by the individual methods for community detection; in details, \textit{Louvain}, \textit{Greedy}, and \textit{Infomap} take $\sim0.17$, $\sim1.89$, and $\sim14.09$ seconds, respectively.

According to our results, the most performing method in terms of modularity is \textit{Louvain} (see Table \ref{tab:lp} and Table \ref{tab:mdlr}); therefore, we defined the latter by default. However, \textit{Greedy} and \textit{Infomap} can be chosen as other options.

In order to evaluate the effectiveness of \textit{MuLaN}, we generated the same local alignment built with \textit{Louvain} by using \textit{Infomap} and \textit{Greedy}, and we analyzed the results.
We selected those algorithms for the best and fast performance to extract community in multilayer network according to literature
\cite{magnani2021community}.

Since, to the best of our knowledge, \textit{MultiLoAl} and its extension \textit{MuLaN} are the only local alignment algorithms of multilayer networks in the literature, we evaluated the effectiveness of \textit{MuLaN}, by comparing the results obtained with default \textit{MuLaN} version, in which Louvain is used as community detection algorithm and other \textit{MuLaN} versions in which are applied \textit{Infomap} and \textit{Greedy} algorithms.

Then, we measured the performance of the alignments built with different versions of \textit{MuLaN} by evaluating the quality of the results.
In particular, we measure the topological quality of alignments and the quality of communities found. 

\subsubsection{Topological quality}
At first, the results are evaluated by analyzing the topological quality.
We recall that intuitively an alignment is of high topological quality if it reconstructs the underlying true node mapping well (when such mapping is known) and if it conserves many edges. For example, for simple networks the ${F-NC}$ (F-score node correctness) is applied to measure the node correctness, and it is defined as $\frac{M\cap
N}{N}$ where $M$ is the set of node pairs that are mapped under the true node mapping and 
$N$ the set of node pairs that are aligned under an alignment $f$.

Otherwise, {NCV-G$S^3$} (high node coverage (NCV) and Generalized $S^3$ (G$S^3$)) is applied to measure the edge correctness, and it is defined as the geometric mean of high node coverage (NCV) and generalized $S^3$ (G$S^3$) measures. NCV is the percentage of nodes from $G_{1}$ and $G_{2}$ that are also in $G'_{1}$ and $G'_{2}$ and G$S^3$measures how well edges are conserved
between $G'_{1}$ and $G'_{2}$ where $G_{1}$ and $G_{2}$ are two graphs and $G'_{1}$ and $G'_{2}$ are subgraphs of $G_1$ and $G_2$ that are induced by the mapping.

In a previous work (\cite{milano2022design}), we extended such measures in the multilayer case, due to, to the best of our knowledge, there are not any other available measures.

In particular, we compute the multilayer ${F-NC_{m}}$ as the average of ${F-NC_{i}}$ estimated for each layer.
Likewise, we compute the multilayer ${NCV-GS^3_m}$ as the average of ${NCV-GS^3_m}$ estimated for each layer.

Finally, we consider the edge correctness for the \textit{inter}-layer edges. Without loss of information, we consider all the \textit{inter}-layer edges as a whole, and we calculate the correctness of all the \textit{inter}-layer edges as ${NCV-GS^3_inter}$.

\begin{figure}[!ht]
\centering
\includegraphics[width=0.8\textwidth]{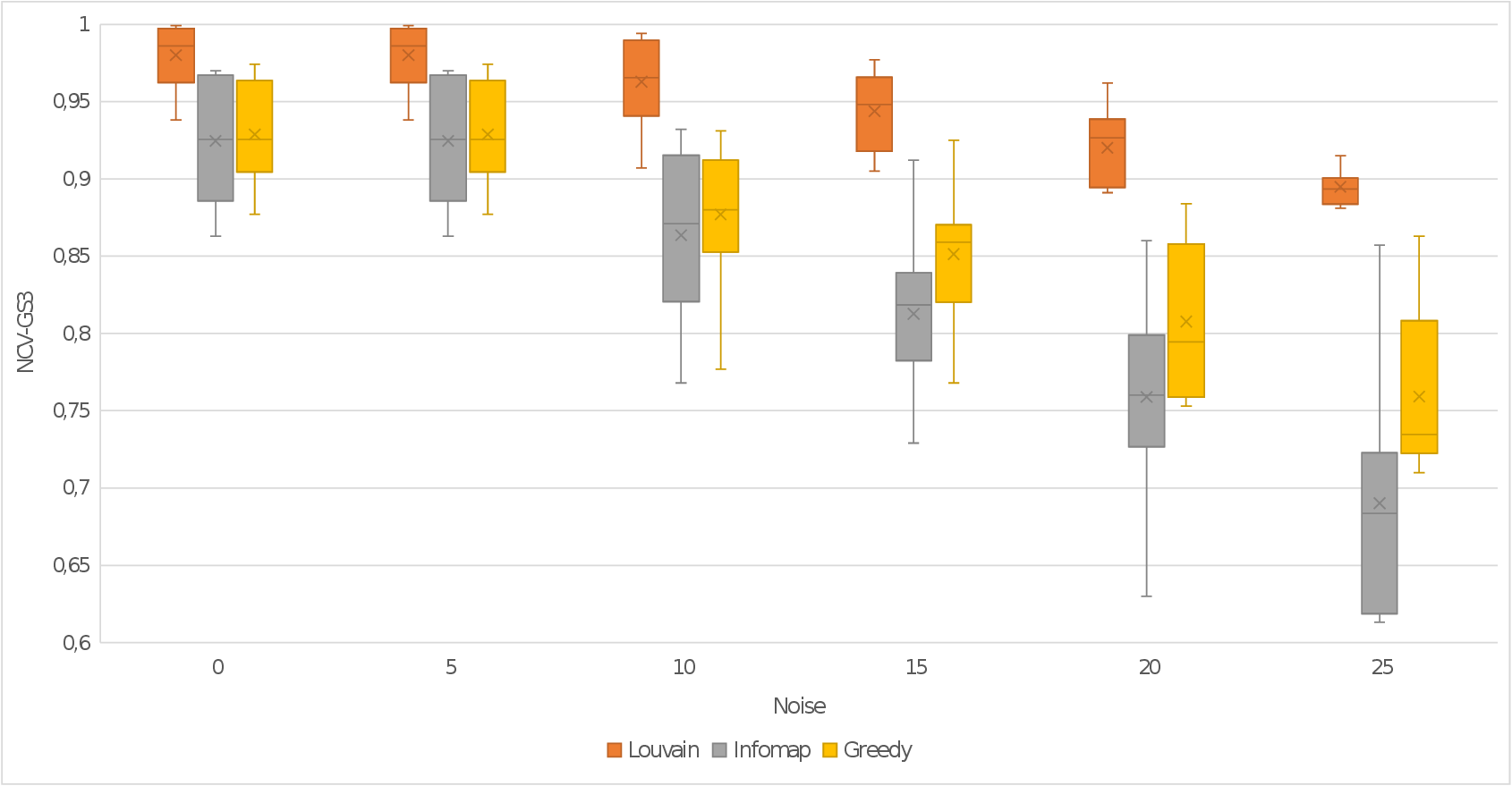}
\caption{NCV-G$S^3$ values computed on \textit{intra}-layer edges for all the local alignments built on synthetic networks by applying \textit{Louvain}, \textit{Infomap}, \textit{Greedy} community detection algorithms.} 
\label{fig:gs3intra}
\end{figure}
\begin{figure}[!ht]
\centering
\includegraphics[width=0.8\textwidth]{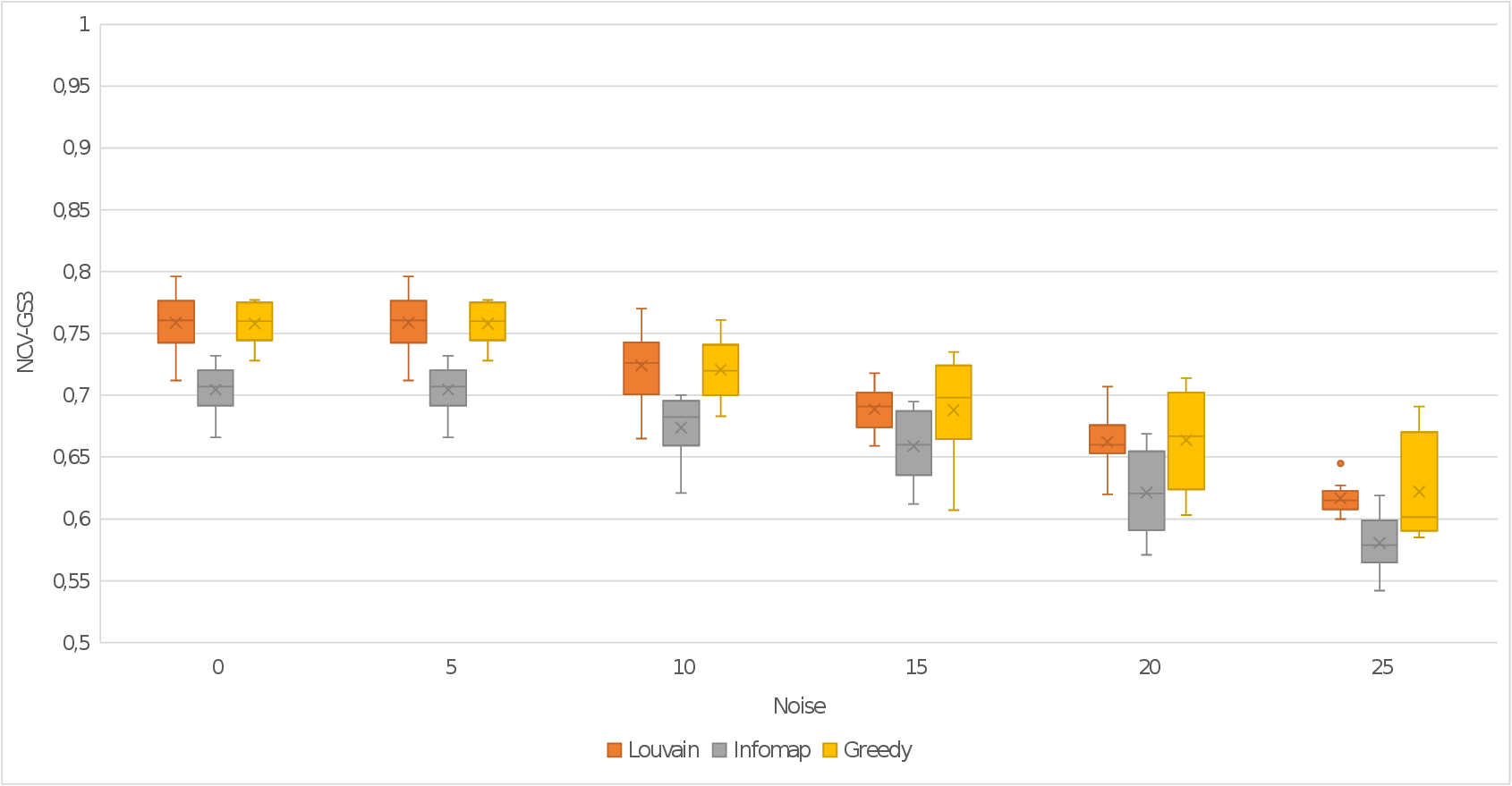}
\caption{NCV-G$S^3$ values computed on \textit{inter}-edges for all the local alignments built on synthetic networks by applying \textit{Louvain}, \textit{Infomap}, \textit{Greedy} community detection algorithms.} 
\label{fig:gs3inter}
\end{figure}
\begin{figure}[!ht]
\centering
\includegraphics[width=0.8\textwidth]{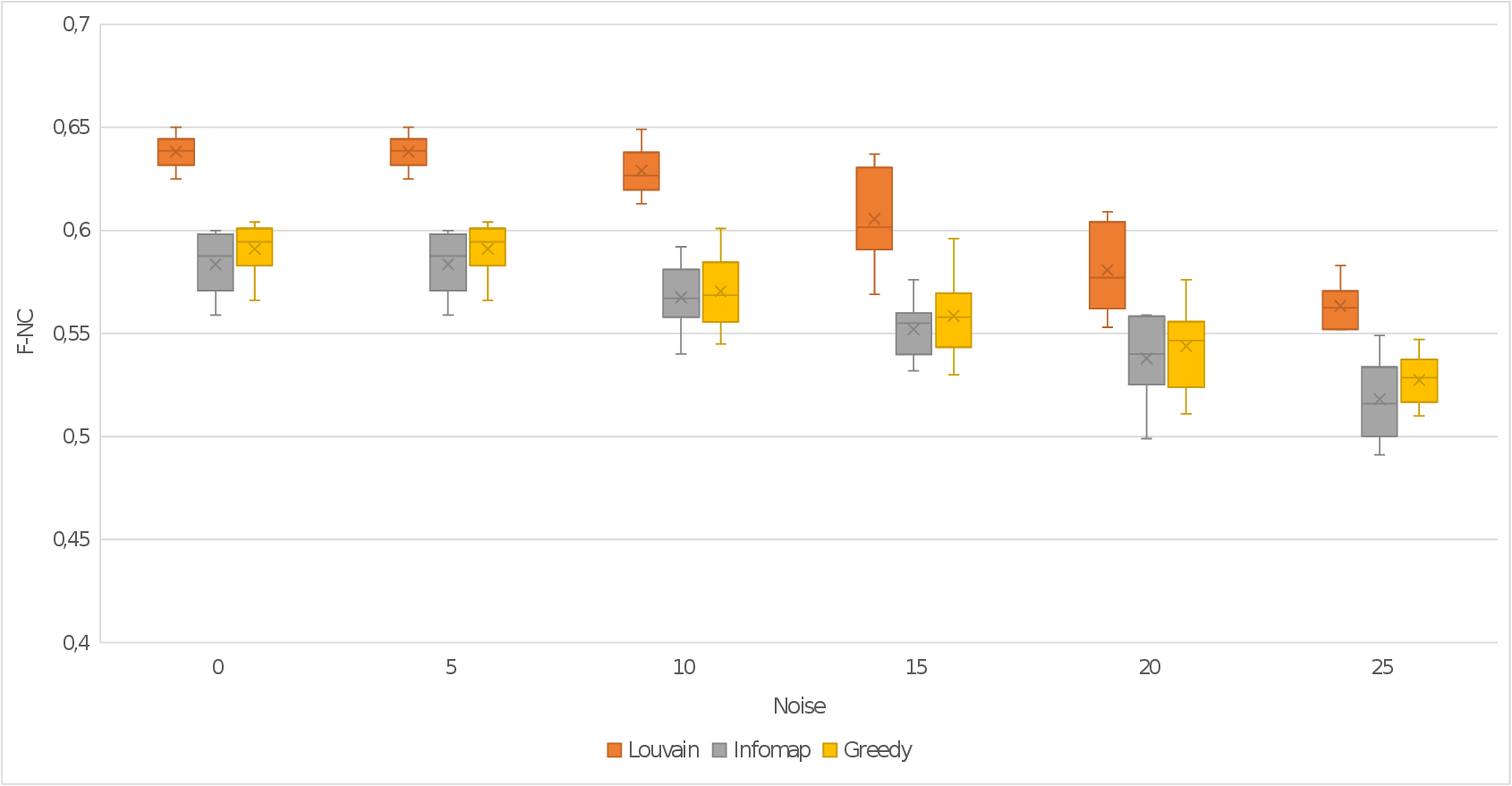}
\caption{F-NC values computed on \textit{intra}-layer edges for all the local alignments built on synthetic networks by applying \textit{Louvain}, \textit{Infomap}, \textit{Greedy} community detection algorithms.} 
\label{fig:fncintra}
\end{figure}
\begin{figure}[!ht]
\centering
\includegraphics[width=0.8\textwidth]{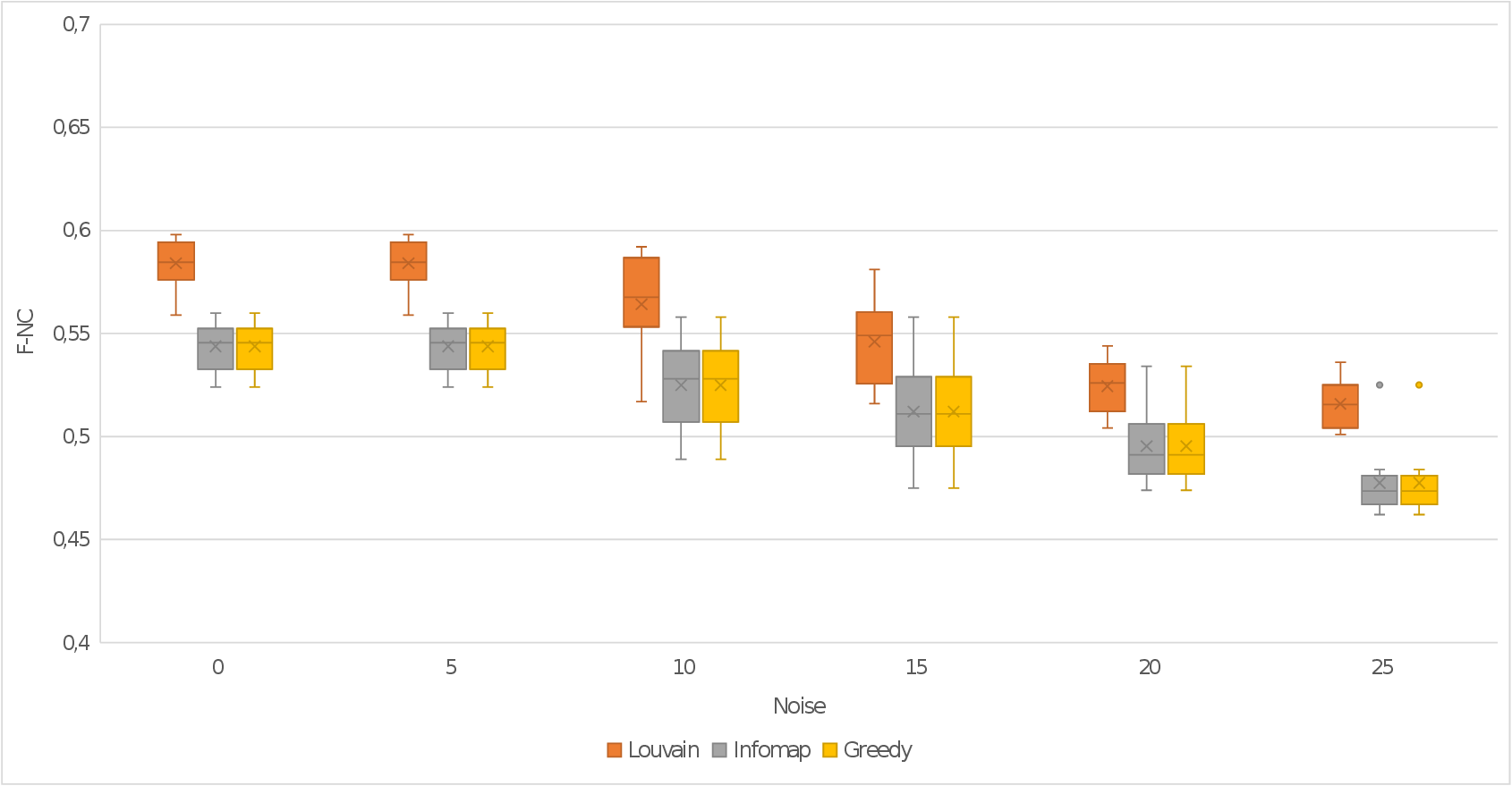}
\caption{F-NC values computed on \textit{inter}-layer edges for all the local alignments built on synthetic networks by applying \textit{Louvain}, \textit{Infomap}, \textit{Greedy} community detection algorithms.} 
\label{fig:fncinter}
\end{figure}
\begin{figure}[!ht]
\centering
\includegraphics[width=0.8\textwidth]{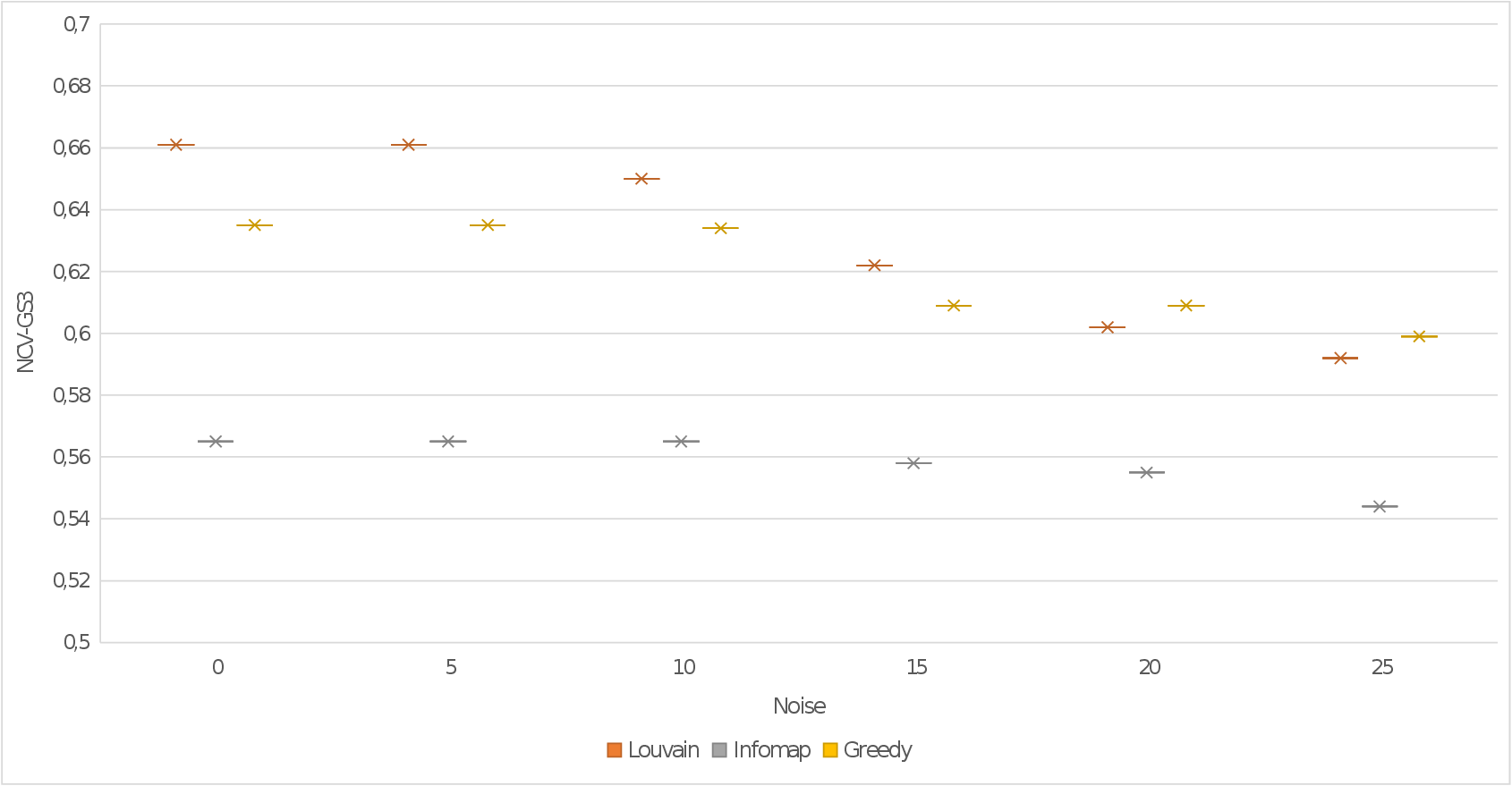}
\caption{NCV-G$S^3$ values computed on \textit{intra}-edges for all the local alignments built on real network by applying \textit{Louvain}, \textit{Infomap}, \textit{Greedy} community detection algorithms.}
\label{fig:Rgs3intra}
\end{figure}
\begin{figure}[!ht]
\centering
\includegraphics[width=0.8\textwidth]{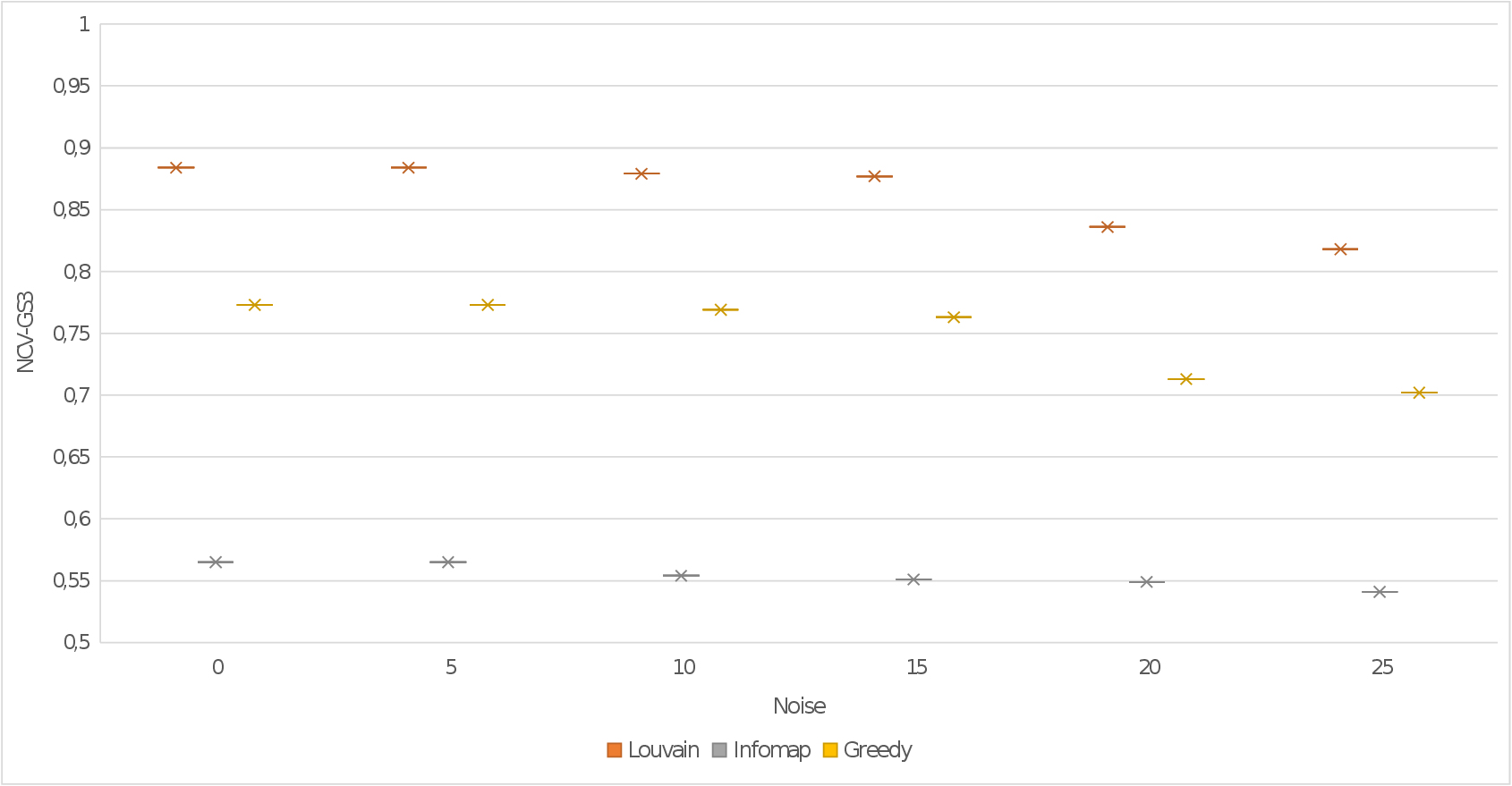}
\caption{NCV-G$S^3$ values computed on \textit{inter}-edges for all the local alignments built on real network by applying \textit{Louvain}, \textit{Infomap}, \textit{Greedy} community detection algorithms.}
\label{fig:Rgs3inter}
\end{figure}
\begin{figure}[!ht]
\centering
\includegraphics[width=0.8\textwidth]{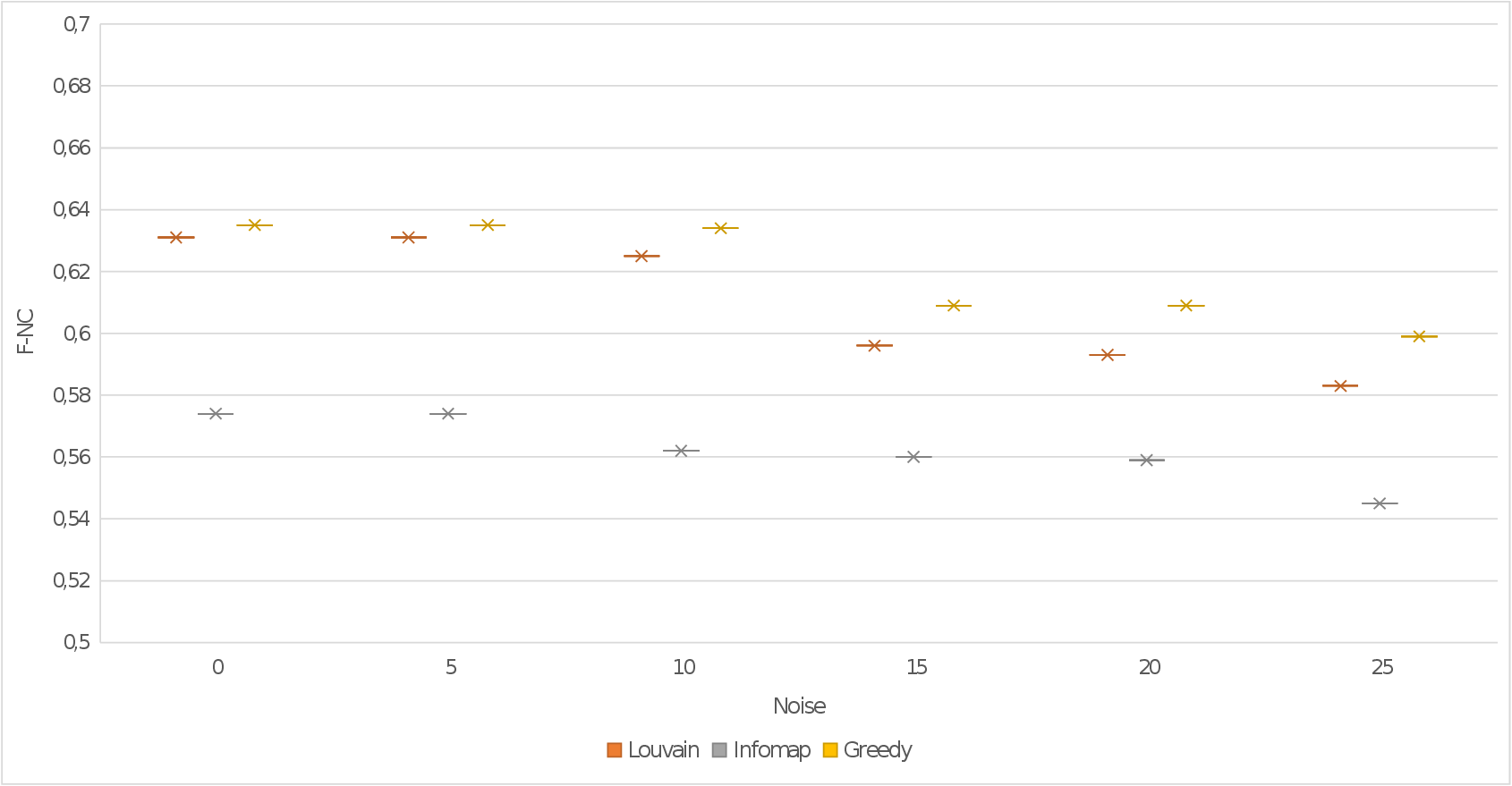}
\caption{F-NC values computed on \textit{intra}-edges for all the local alignments built on real network by applying \textit{Louvain}, \textit{Infomap}, \textit{Greedy} community detection algorithms.} 
\label{fig:Rfncintra}
\end{figure}
\begin{figure}[!ht]
\centering
\includegraphics[width=0.8\textwidth]{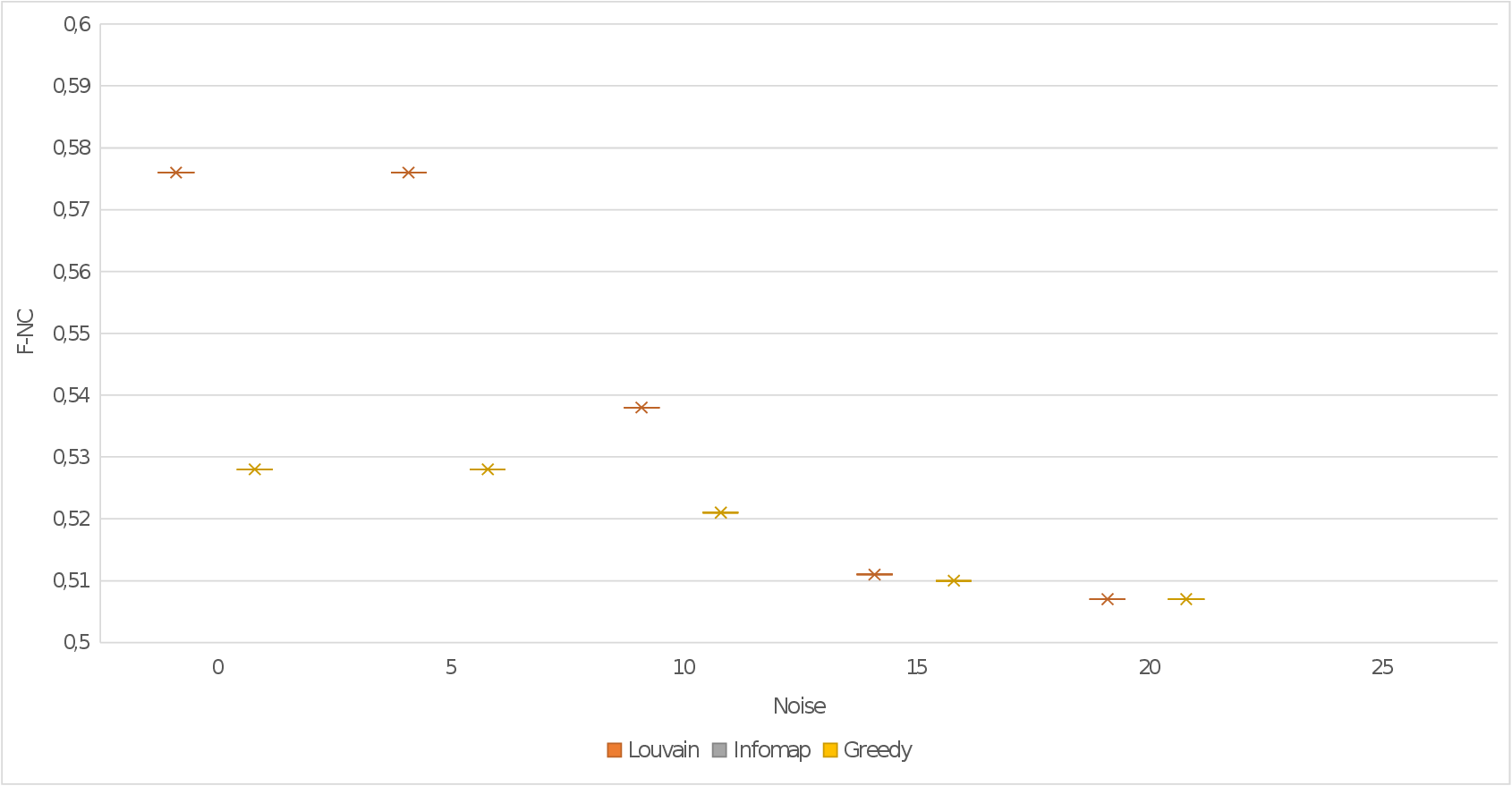}
\caption{F-NC values computed on \textit{inter}-edges for all the local alignments built on real network by applying \textit{Louvain}, \textit{Infomap}, \textit{Greedy} community detection algorithms.} 
\label{fig:Rfncinter}
\end{figure}

We computed multilayer ${NCV-GS^3_m}$ and multilayer ${F-NC_{m}}$ measures for all alignments built for each synthetic network and for real network by considering the \textit{intra}-layer and \textit{inter}-layer. 
Figure \ref{fig:gs3intra}, Figure \ref{fig:gs3inter}, Figure \ref{fig:fncintra}, Figure \ref{fig:fncinter} report the results for synthetic multilayer networks, whereas Figure \ref{fig:Rgs3intra}, Figure \ref{fig:Rgs3inter}, Figure \ref{fig:Rfncintra}, Figure \ref{fig:Rfncinter} report the results for real multilayer network.  The Tables related to ${NCV-GS^3_m}$ and multilayer ${F-NC_{m}}$ measures for all alignments built for each synthetic network and for real network by considering the \textit{intra}-layer and \textit{inter}-layer are reported in appendix.

By analyzing the results, it is possible to notice that multilayer ${NCV-GS^3_m}$ and multilayer ${F-NC_{m}}$ values decrease by increasing noise level from 5 \% to 25 \%. 
The reduction of quality of the alignment is detected for different \textit{MuLaN} versions applied on both synthetic network and real network.
Otherwise, the results show that the quality of the alignment is greater when \textit{Louvain} is applied to extract communities.

\subsubsection{Community Quality}
We also evaluated the quality of mined communities by considering i) the number of extracted communities and ii) the strength of the community structure by using the modularity \cite{newman2006modularity} as a community quality metric.
In detail, modularity is a measure related to the network structure that detects the density of connections within a module. The modularity can be either positive or negative, and positive values are indicative of the presence of community structure. Thus, a network with a high modularity score will have many connections within a community \cite{newman2006modularity}.

Thus, we first estimated the number of mined communities with \textit{Louvain}, \textit{Infomap} and \textit{Greedy}, for each local alignment built on synthetic and real networks, then, we computed the modularity of the extracted communities. 

Table \ref{tab:lp} and Table \ref{tab:mdlr} report the results obtained with different versions of the \textit{MuLaN} algorithm for synthetic and real networks.
In particular, for each network, the Tables report the number of the extracted communities and the modularity value for each community.
The result shows that, all the community detection algorithms are able to mine more communities by aligning the original network with itself.
Otherwise, the number of communities decreases on the alignment built with the synthetic counterparts with 5 \%, 10\%, 15\%, 20\% and 25\% of added noise.
Furthermore, it is possible to notice that 
\textit{Louvain} is able to identify a higher number of communities in the multilayer network with respect to other community detection algorithms.

By analyzing the results, it is possible to see that the higher modularity values are obtained by applying \textit{Louvain} as a community extraction algorithm.
This means that \textit{Louvain} is able to extract communities that reveal the relationships among nodes in various layers, with respect to other community detection algorithms.

\begin{table*}[!ht]
\centering
\scriptsize
\caption{The Table reports for each synthetic network: the level of noise, number of Communities extracted and Modularity values by applying \textit{Louvain} , \textit{Infomap} , \textit{Greedy} algorithms respectively.}
\label{tab:lp}
\scalebox{0.53}{
\begin{tabular}{|p{2cm}|p{2cm}|p{2cm}|p{2cm}|p{2cm}|p{2cm}|p{2cm}|p{2cm}|}
\hline

\textbf{Network } & \textbf{Noise} & \textbf{$\#$ Communities with \textit{Louvain}}& \textbf{Modularity with  \textit{Louvain} } & \textbf{$\#$ Communities with \textit{Infomap} } & \textbf{Modularity with \textit{Infomap} } & \textbf{$\#$ Communities with \textit{Greedy} } & \textbf{Modularity with \textit{Greedy} } \\ \hline
\multirow{6}{*}{N1} &	0	&	37	&	0,844	&	28	&	0,839	&	38	&	0,844	\\	
&	5	&	40	&	0,850	&	26	&	0,845	&	42	&	0,851	\\	
&	10	&	40	&	0,857	&	29	&	0,855	&	44	&	0,857	\\	
&	15	&	40	&	0,860	&	30	&	0,857	&	45	&	0,861	\\	
&	20	&	42	&	0,872	&	28	&	0,867	&	46	&	0,873	\\	
&	25	&	45	&	0,872	&	29	&	0,869	&	46	&	0,873	\\	\hline
\multirow{6}{*}{N2} &	0	&	39	&	0,840	&	27	&	0,836	&	40	&	0,841	\\	
&	5	&	40	&	0,848	&	28	&	0,844	&	42	&	0,849	\\	
&	10	&	40	&	0,859	&	32	&	0,857	&	42	&	0,859	\\	
&	15	&	41	&	0,861	&	30	&	0,858	&	43	&	0,859	\\	
&	20	&	41	&	0,870	&	27	&	0,856	&	46	&	0,861	\\	
&	25	&	42	&	0,859	&	30	&	0,866	&	46	&	0,871	\\	\hline
\multirow{6}{*}{N3} &	0	&	38	&	0,843	&	28	&	0,840	&	37	&	0,843	\\	
&	5	&	39	&	0,846	&	26	&	0,844	&	40	&	0,848	\\	
&	10	&	37	&	0,868	&	31	&	0,854	&	42	&	0,856	\\	
&	15	&	41	&	0,857	&	27	&	0,856	&	47	&	0,859	\\	
&	20	&	43	&	0,859	&	28	&	0,865	&	44	&	0,867	\\	
 &	25	&	46	&	0,877	&	26	&	0,870	&	48	&	0,877	\\	\hline
\multirow{6}{*}{N4} &	0	&	41	&	0,843	&	29	&	0,839	&	41	&	0,843	\\	
&	5	&	39	&	0,851	&	31	&	0,850	&	43	&	0,852	\\	
&	10	&	42	&	0,859	&	29	&	0,856	&	40	&	0,859	\\	
&	15	&	42	&	0,870	&	30	&	0,854	&	44	&	0,859	\\	
&	20	&	43	&	0,867	&	32	&	0,862	&	50	&	0,867	\\	
&	25	&	44	&	0,858	&	29	&	0,866	&	44	&	0,870	\\	\hline
\multirow{6}{*}{N5} &	0	&	38	&	0,842	&	29	&	0,839	&	41	&	0,841	\\	
&	5	&	41	&	0,867	&	29	&	0,844	&	42	&	0,849	\\	
&	10	&	42	&	0,849	&	30	&	0,852	&	43	&	0,854	\\	
&	15	&	42	&	0,863	&	31	&	0,856	&	45	&	0,863	\\	
&	20	&	42	&	0,874	&	34	&	0,863	&	47	&	0,867	\\	
&	25	&	45	&	0,854	&	30	&	0,871	&	48	&	0,874	\\	\hline
\multirow{6}{*}{N6} &	0	&	40	&	0,841	&	27	&	27,000	&	39	&	0,841	\\	
&	5	&	40	&	0,845	&	29	&	0,842	&	40	&	0,845	\\	
&	10	&	42	&	0,856	&	27	&	0,842	&	46	&	0,856	\\	
&	15	&	45	&	0,889	&	28	&	0,867	&	46	&	0,872	\\	
&	20	&	46	&	0,872	&	35	&	0,887	&	50	&	0,888	\\	
&	25	&	60	&	0,900	&	38	&	0,894	&	62	&	0,900	\\	\hline
\multirow{6}{*}{N7} &	0	&	37	&	0,845	&	29	&	0,841	&	39	&	0,844	\\	
&	5	&	38	&	0,854	&	28	&	0,850	&	39	&	0,854	\\	
&	10	&	43	&	0,867	&	30	&	0,859	&	43	&	0,867	\\	
&	15	&	49	&	0,880	&	31	&	0,874	&	50	&	0,881	\\	
&	20	&	52	&	0,904	&	35	&	0,889	&	55	&	0,893	\\	
&	25	&	53	&	0,892	&	37	&	0,900	&	59	&	0,904	\\	\hline
\multirow{6}{*}{N8} &	0	&	40	&	0,843	&	27	&	0,840	&	41	&	0,844	\\	
&	5	&	40	&	0,849	&	27	&	0,848	&	40	&	0,849	\\	
&	10	&	39	&	0,861	&	28	&	0,856	&	43	&	0,861	\\	
&	15	&	42	&	0,878	&	28	&	0,874	&	50	&	0,878	\\	
&	20	&	46	&	0,887	&	29	&	0,879	&	58	&	0,888	\\	
&	25	&	49	&	0,902	&	37	&	0,897	&	58	&	0,903	\\	\hline
\multirow{6}{*}{N9} &	0	&	36	&	0,852	&	24	&	0,839	&	40	&	0,844	\\	
&	5	&	40	&	0,852	&	28	&	0,850	&	40	&	0,851	\\	
&	10	&	39	&	0,861	&	30	&	0,858	&	44	&	0,861	\\	
&	15	&	42	&	0,874	&	33	&	0,872	&	47	&	0,874	\\	
&	20	&	44	&	0,887	&	32	&	0,883	&	47	&	0,886	\\	
&	25	&	52	&	0,900	&	35	&	0,895	&	52	&	0,901	\\	\hline
\multirow{6}{*}{N10} &	0	&	38	&	0,840	&	29	&	0,838	&	41	&	0,841	\\	
&	5	&	37	&	0,849	&	26	&	0,847	&	41	&	0,849	\\	
&	10	&	39	&	0,861	&	28	&	0,858	&	42	&	0,862	\\	
&	15	&	44	&	0,877	&	32	&	0,874	&	48	&	0,876	\\	
&	20	&	52	&	0,890	&	33	&	0,884	&	56	&	0,890	\\	
&	25	&	53	&	0,897	&	36	&	0,892	&	55	&	0,898	\\	\hline
\hline
\end{tabular}}
\end{table*}

\begin{table*}[!ht]
\centering
\scriptsize
\caption{The Table reports for the real network: the level of noise, number of Communities extracted and Modularity values by applying \textit{Louvain} algorithm, \textit{Infomap} algorithm, \textit{Greedy} algorithm.}
\label{tab:mdlr}
\scalebox{0.53}{
\begin{tabular}{|p{2cm}|p{2cm}|p{2cm}|p{2cm}|p{2cm}|p{2cm}|p{2cm}|p{2cm}|}
\hline

\textbf{Network } & \textbf{Noise} & \textbf{$\#$ Communities with \textit{Louvain} }& \textbf{ Modularity with Generalized \textit{Louvain} } & \textbf{$\#$ Communities with \textit{Infomap} } & \textbf{Modularity with \textit{Infomap} } & \textbf{$\#$ Communities with \textit{Greedy} } & \textbf{Modularity with \textit{Greedy} } \\ \hline
\multirow{5}{*}{N1} &	0	&	267	&	0.335	&	50	&	0.036	&	302	&	0.270	\\	
&	5	&	268	&	0.331	&	49	&	0.034	&	301	&	0.268	\\	
&	10	&	266	&	0.331	&	60	&	0.036	&	297	&	0.269	\\	
&	15	&	260	&	0.331	&	46	&	0.035	&	340	&	0.250	\\	
&	20	&	260	&	0.329	&	52	&	0.034	&	312	&	0.265	\\	
&	25	&	263	&	0.327	&	57	&	0.031	&	307	&	0.260	\\	\hline
\hline
\end{tabular}}
\end{table*}

\subsubsection{Functional Quality Evaluation}
Finally, we evaluate the functional quality of the results by considering the biological relevance of the extracted communities.
For classical networks, the biological relevance evaluation consists in evaluating whether groups of related entities have a similar biological role or share functions. 
Since, \textit{MuLaN} constructs the local alignment that includes associations between different entities, i.e. disease-drug, we conducted the evaluation differently.
In particular, we evaluated the functional quality of local alignment, considering whether our algorithm is able to extract drug-disease associations still unobserved.

We recall that, drug-disease associations consist of the cases in which drugs affect disease. Thus, in research field, drug-disease associations may bring relevant information in the drug discovery, for example, the detection of convenient relations between drugs and diseases.
The literature contain a large set of drug-disease associations already identified, however many associations are unobserved or not detected.
In the rest of the paper, we refer to unobserved or not detected drug-disease associations as candidate drug-disease associations.

Starting from this consideration, our goal is to demonstrate that \textit{MuLaN} is able to build a local alignment containing candidate drug-disease associations, and it is able to extract knowledge about the aligned multilayer networks.

For this aim, we consider the local alignment constructed by aligning the real network with its noisy counterpart at 25\% (because it is the most different from the original network) and we analyze the drug-disease associations forming the extracted communities.

To evaluate whether the associations contained in the communities are known in the literature or still unobserved drug-disease associations, we used SCMFDD \cite{zhang2018predicting}, a similarity constrained matrix factorization method for the drug-disease association prediction.
SCMFDD reveals unobserved drug-disease associations by using drug features, disease semantic information and known associations incorporated into the matrix factorization frame (see \cite{zhang2018predicting} for complete details).

SCMFDD computes: i) the drug-drug similarities by using Jaccard index \cite{real1996probabilistic} taking into account diverse drug features (such as, targets, enzymes, pathways and drug-drug interactions), ii) the disease-disease semantic similarity by using MeSH information \cite{xuan2013prediction}, iii) two low-rank feature matrices by applying a factorization on observed drug-disease associations matrix.
Then, it uses an objective function based on Newton's method \cite{berahas2019derivative} to compute a score representing the probability that a drug and a disease have an association (see \cite{zhang2018predicting} for complete details on SCMFDD).
Thus, we applied SCMFDD to analyze each couple of drug-disease associations forming the community extracted with \textit{MuLaN}.
The results show that the extracted communities contain 46,498 candidate drug-disease associations. Among these, 711 drug-disease associations present a score greater than $0.5$.
We reported candidate drug-disease associations with score equal to 1 in Table \ref{tab:my_table}. The complete detected associations are available at \url{https://github.com/pietrocinaglia/mulan}).

Furthermore, we manually evaluated the candidate drug-disease associations by searching literature evidences.
For example, by considering the drug-disease association Fenofibrate-Cholestasis, the works in \cite{dai2017inhibition, ghonem2015fibrates} report the effect of fenofibrate against cholestasis. 
In \cite{ghonem2015fibrates}, Ghonem et al. discuss the effectiveness and well tolerability of Orlistat treatment for hyperglycemia. Also, \cite{yonkof2020successful} presents the successful use of rifampin in a patient with Stevens-Johnson syndrome. We list the literature evidence of the top 10 detected drugs-disease associations in Table \ref{tab:my_table2}.

In conclusion, the results demonstrate that our algorithm is capable of discovering candidate drug-disease associations and, consequently, of extracting new knowledge from multilayer networks. 

\begin{table}[!ht]
\centering
\caption{The Table reports an example of novel drug-disease associations contained in the local alignment built with \textit{MuLaN}. }
\label{tab:my_table}
\scalebox{0.7}{
\begin{tabular}{lllll}
\hline
Drug ID	&	Drug Name	&	Disease ID	&	Disease Name	&	Score	 \\ \hline
DB01039	&	Fenofibrate	&	DOID:13580	&	Cholestasis	&	1	 \\ \hline
DB01083	&	Orlistat	&	DOID:4195	&	Hyperglycemia	&	1	 \\ \hline
DB01220	&	Rifaximin	&	DOID:0050426	&	Stevens-Johnson syndrome	&	1	 \\ \hline
DB00951	&	Isoniazid	&	DOID:2355	&	Anemia	&	1	 \\ \hline
DB00570	&	Vinblastine	&	DOID:2355	&	Anemia	&	1	 \\ \hline
DB01241	&	Gemfibrozil	&	DOID:13580	&	Cholestasis	&	1	 \\ \hline
DB01602	&	Bacampicillin	&	DOID:0050426	&	Stevens-Johnson syndrome	&	1	 \\ \hline
DB01042	&	Melphalan	&	DOID:987	&	Alopecia	&	1	 \\ \hline
DB00535	&	Cefdinir	&	DOID:13580	&	Cholestasis	&	1	 \\ \hline
DB01005	&	Hydroxyurea	&	DOID:1227	&	Neutropenia	&	1	 \\ \hline
DB01593	&	Zinc	&	DOID:2355	&	Anemia	&	1	 \\ \hline
DB00993	&	Azathioprine	&	DOID:1227	&	Neutropenia	&	1	 \\ \hline
DB02659	&	Cholic Acid	&	DOID:13580	&	Cholestasis	&	1	 \\ \hline
DB01103	&	Quinacrine	&	DOID:13580	&	Cholestasis	&	1	 \\ \hline
DB00970	&	Dactinomycin	&	DOID:1227	&	Neutropenia	&	1	 \\ \hline
DB00586	&	Diclofenac	&	DOID:2355	&	Anemia	&	1	 \\ \hline
DB00290	&	Bleomycin	&	DOID:2355	&	Anemia	&	1	 \\ \hline
DB00963	&	Bromfenac	&	DOID:13580	&	Cholestasis	&	1	 \\ \hline
DB01137	&	Levofloxacin	&	DOID:13580	&	Cholestasis	&	1	 \\ \hline
DB01262	&	Decitabine	&	DOID:299	&	Adenocarcinoma	&	1	 \\ \hline
DB01008	&	Busulfan	&	DOID:2355	&	Anemia	&	1	 \\ \hline
DB00549	&	Zafirlukast	&	DOID:13580	&	Cholestasis	&	1	 \\ \hline
DB00763	&	Methimazole	&	DOID:0050426	&	Stevens-Johnson syndrome	&	1	 \\ \hline
DB00591	&	Fluocinolone Ace	&	DOID:1555	&	Urticaria	&	1	 \\ \hline
DB01022	&	Phylloquinone	&	DOID:1555	&	Urticaria	&	1	 \\ \hline
DB00271	&	Diatrizoate	&	DOID:10763	&	Hypertension	&	1	 \\ \hline
DB01143	&	Amifostine	&	DOID:615	&	Leukopenia	&	1	 \\ \hline
DB01015	&	Sulfamethoxazole	&	DOID:1555	&	Urticaria	&	1	 \\ \hline
DB00984	&	Nandrolone phenp	&	DOID:10763	&	Hypertension	&	1	 \\ \hline
DB00296	&	Ropivacaine	&	DOID:10763	&	Hypertension	&	1	 \\ \hline
DB00205	&	Pyrimethamine	&	DOID:1588	&	Thrombocytopenia	&	1	 \\ \hline
DB01254	&	Dasatinib	&	DOID:13250	&	Diarrhea	&	1	 \\ \hline
DB00291	&	Chlorambucil	&	DOID:1588	&	Thrombocytopenia	&	1	 \\ \hline
DB00293	&	Raltitrexed	&	DOID:615	&	Leukopenia	&	1	 \\ \hline
DB01406	&	Danazol	&	DOID:2237	&	Hepatitis	&	1	 \\ \hline
DB00456	&	Cefalotin	&	DOID:1555	&	Urticaria	&	1	 \\ \hline
DB01138	&	Sulfinpyrazone	&	DOID:576	&	Proteinuria	&	1	 \\ \hline
DB00480	&	Lenalidomide	&	DOID:615	&	Leukopenia	&	1	 \\ \hline
DB01161	&	Chloroprocaine	&	DOID:10763	&	Hypertension	&	1	 \\ \hline
DB00287	&	Travoprost	&	DOID:10763	&	Hypertension	&	1	 \\ \hline
DB00121	&	Biotin	&	DOID:10763	&	Hypertension	&	1	 \\ \hline
DB00491	&	Miglitol	&	DOID:10763	&	Hypertension	&	1	 \\ \hline
DB00459	&	Acitretin	&	DOID:2237	&	Hepatitis	&	1	 \\ \hline
DB00905	&	Bimatoprost	&	DOID:10763	&	Hypertension	&	1	 \\ \hline
DB01288	&	Fenoterol	&	DOID:10763	&	Hypertension	&	1	 \\ \hline
DB04272	&	Citric acid	&	DOID:10763	&	Hypertension	&	1	 \\ \hline
DB01086	&	Benzocaine	&	DOID:1555	&	Urticaria	&	1	 \\ \hline
\end{tabular}
 }
\end{table}

\begin{table}[!ht]
\centering
\caption{Top 10 drug-disease associations contained in the local alignment built with \textit{MuLaN}. }
\label{tab:my_table2}
\scalebox{0.7}{
\begin{tabular}{lll}
\hline
Drug Name		&	Disease Name	&	Evidence	 \\ \hline
Fenofibrate	&	Cholestasis	&	\cite{dai2017inhibition, ghonem2015fibrates}	 \\ \hline
Orlistat	&	Hyperglycemia	&	\cite{ganjayi2023quercetin}		 \\ \hline
Rifaximin	&	Stevens-Johnson syndrome	&	\cite{yonkof2020successful}		 \\ \hline
Isoniazid	&	Anemia	&	\cite{romano2022next}		 \\ \hline
Vinblastine	&	Anemia	&	\cite{vagrecha2022extracorpuscular}	 \\ \hline
Gemfibrozil	&	Cholestasis	&	\cite{yang2023integration}		 \\ \hline
Bacampicillin	&	Stevens-Johnson syndrome	&	\cite{romano2022next}		 \\ \hline
Melphalan	&	Alopecia	&	\cite{gurevich2022comparison}		 \\ \hline
Cefdinir	&	Cholestasis	&	\cite{phillips2022beta}		 \\ \hline
Hydroxyurea	&	Neutropenia	&	\cite{akram2022efficacy}		 \\ \hline\hline
\end{tabular}
 }
\end{table}

\subsubsection{\textit{MuLaN} vs \textit{MultiLoAl}}
To clarify the need for developing  an extended version of \textit{MultiLoAl}, in this section, we present the comparison of \textit{MuLaN} with \textit{MultiLoAl}.The aim is to demonstrate that \textit{MuLaN} reports the best performance to analyze multilayer networks with respect to previous version \textit{MultiLoAl}. The dataset that we used for the comparison consists of ten synthetic multilayer networks and one real multilayer network used for \textit{MuLaN} evaluation.
Thus, we built the alignment of a network with respect to itself, and considering the alignment of a network with respect to an altered version by applying \textit{MultiLoAl}. Our experimentation shows that \textit{MultiLoAl} builds the alignment between two synthetic networks, having $2000$ nodes and $2298$ edges each one, in $\sim120$ minutes, against the $\sim1.42$ seconds required by \textit{MuLaN} for the same operation. 

Results show clearly that \textit{MuLaN} outperforms \textit{MultiLoAl}, in terms of memory efficiency and runtime.

\section{Conclusions}
\label{section:conclusions}
Multilayer networks are a powerful tool for the modelling of complex data in biology and medicine.
In this work, we focused on the problem of 
analysis of multilayer networks through the comparison of their internal structure, by applying Network Alignment algorithms.
Since classical network alignment algorithms for simple networks do not perform well on multilayer networks, we presented a deep extension of a previously developed network alignment algorithm for multilayer network, named, \textit{MuLaN} (Local Alignment Algorithm for Multilayer Networks). 
We tested it on ten synthetic multilayer networks and on a real multilayer network, and we evaluated the quality of results. Results confirm that \textit{MuLaN} is able to build alignment with high topological quality. According to these analyses, we demonstrated the \textit{MuLaN} methodology is suitable for the comparison and hence, the analysis of multilayer networks.

\section{Competing interests}
No competing interest is declared.

\section{Author contributions statement}
Conceptualization, PHG. and MM.; methodology, PHG, PC, and MM; software, PC, MM; data curation, MM and PC; writing--original draft preparation, MM, PC, PHG, and MC.; writing--review and editing, MM, PC, PHG and MC.; funding acquisition, MC. All authors have read and agreed to the published version of the manuscript.

\section{Funding}
This work was funded by the Next Generation EU - Italian NRRP, Mission 4, Component 2, Investment 1.5, call for the creation and strengthening of 'Innovation Ecosystems', building 'Territorial R\&D Leaders' (Directorial Decree n. 2021/3277) - project Tech4You - Technologies for climate change adaptation and quality of life improvement, n. ECS0000009. This work reflects only the authors' views and opinions, neither the Ministry for University and Research nor the European Commission can be considered responsible for them. 

\bibliographystyle{elsarticle-num} 
\bibliography{biblio}

\clearpage
\appendix

\section{Sample Appendix Section}
\label{sec:sample:appendix}
This appendix present the process in which \textit{MuLaN} builds the MAG and results of the computation of ${NCV-GS^3_m}$ and ${F-NC_{m}}$ measures for all alignments built for each synthetic network and for real network by considering the \textit{intra}-layer and \textit{inter}-layer.

\subsection{Building of the Alignment Graph}
\begin{figure}[!ht]
\centering
 \includegraphics[width=0.8\textwidth]{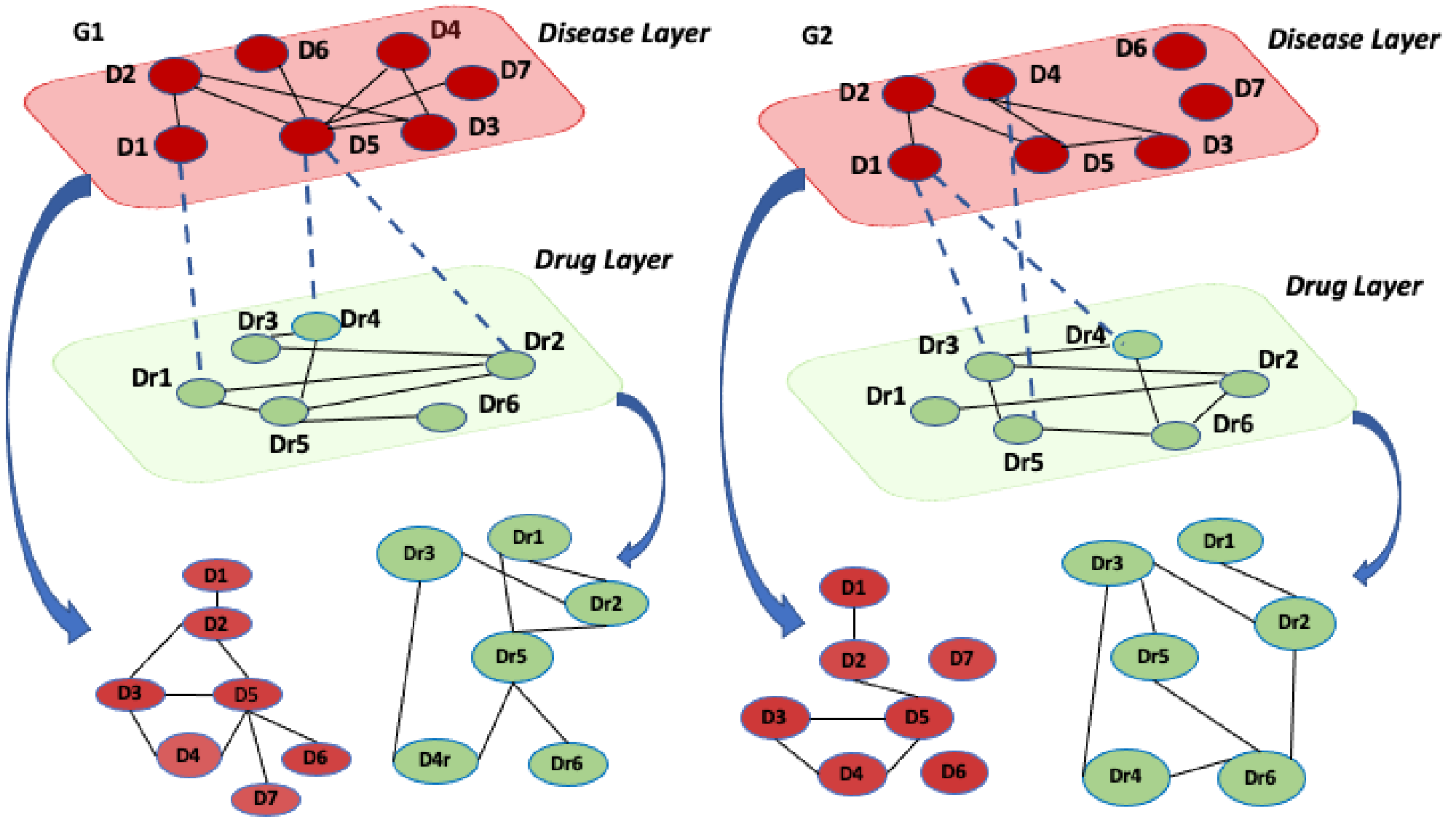}
 \caption{\textit{MuLaN} separates the input networks according to the layer type.}
\label{fig:diap2}
\end{figure}

\begin{figure}[!ht]
\centering
\includegraphics[width=0.8\textwidth]{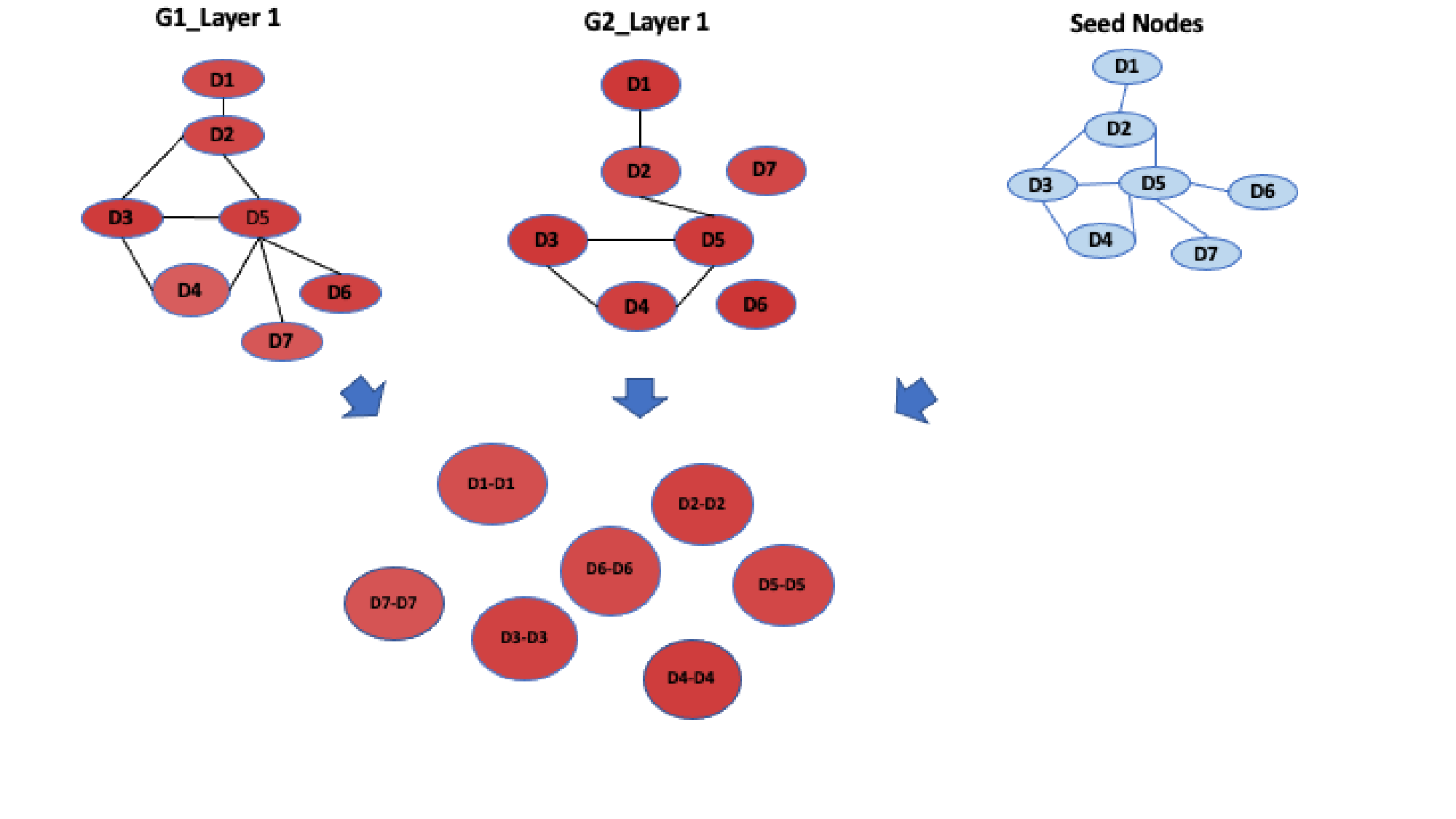}
\caption{Building of alignment graph for the disease layer of the two input multilayer networks: Node Definition. \textit{MuLaN} takes the two networks in a layer and a subset of node pairs matched according to a similarity function and starts to build the alignment graph. In this step, the algorithm defines the nodes of the alignment graph represented by the pair matched nodes.}
\label{fig:diap7}
\end{figure}

It receives as input two multilayer networks ($M_1$ and $M_2$) having the same number of layers $n$ and  a multiset of seed nodes representing similarity between the node of each graph of the same layer $k$ of the two networks.
We assume the existence of a bijective correlation among layers of the networks, so the layer $k$ of $M_1$ correspond to the layer $k$ of $M_2$.  Starting from $M_1$ and $M_2$ and the multiset of seed nodes, in the step 1.a, \textit{MuLaN} analyzes each pair of $k$ layers in  $M_1$  and $M_2$ independently, see Figure \ref{fig:diap2}.  For each layer, \textit{MuLaN} constructs an alignment graph on the basis of the strategy suggested in L-HetNetAligner. \cite{milano2020hetnetaligner}. 
At this stage, \textit{MuLaN} creates the nodes of the alignment graph.
Each node $v_{al}\in V_{al}$ represents a match of nodes of the input graphs, so $V_{al} \subseteq V_{1} \times V_{2}$.
Thus, the choice of node pairs is conducted by the input similarity relationships. Consequently, each node is matched with the most similar node of the other network on the same layer; and each node of the alignment graph represents a pair of similarity among nodes from the input networks, see Figure \ref{fig:diap7}. 

\begin{figure}[!ht]
\centering
\includegraphics[width=0.8\textwidth]{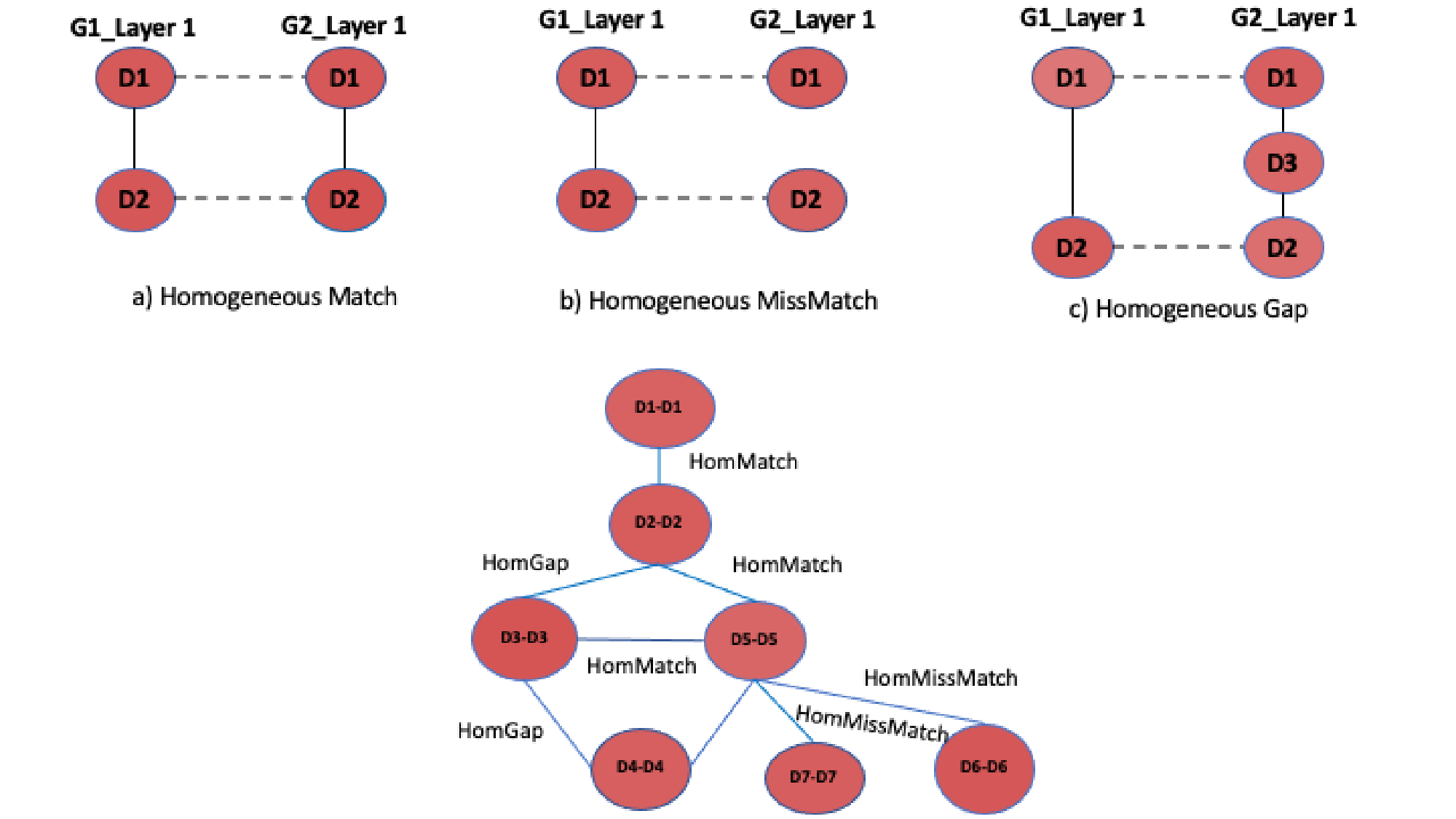}
\caption{Example of homogeneous match, homogeneous mismatch and homogeneous gap and building of Alignment Graph.}
\label{fig:diap4}
\end{figure}

When all nodes have been inserted into the alignment graph, \textit{MuLaN} adds the edges of the alignment graphs. For each pair of nodes, \textit{MuLaN} analyzes the two input graphs, and it inserts and weights the edges considering three conditions: match, mismatch and gap.
In the following, we present each condition.
So, let us consider the nodes of the alignment graphs; in particular, let us analyze the pair of nodes $(D1-D1)$ and $(D2-D2)$ in Figure \ref{fig:diap6}. To determine the presence of an edge, we  we must consider the edges  $(D1,D2) \in G_1$ network and the edges $(D1,D2) \in G_2$ network.
If $G_1$, and $G_2$ contains these nodes, and the nodes are adjacent, we identify this condition  as \textbf{match}, that, for convenience, we call \textbf{homogeneous match}, since the nodes of the two networks are of the same type i.e. disease, (see Figure \ref{fig:diap4} (a)).

Let us consider $\Delta=2$ as node distance, i.e. the length of the shortest connecting path threshold to discriminate between gaps and mismatches.
Let us consider the pair of nodes  $(D1-D1)$ and $(D2-D2)$ (Figure \ref{fig:diap4} (b)).
$G_1$ contains the edges  $(D1,D2)$, whereas the nodes  $(D1-D1)$ are disconnected in $G_2$.  We identify this condition  as \textbf{mismatch} that we call \textbf{homogeneous mismatch}.
Let us consider the pair of nodes  $(D1-D1)$ and $(D2-D2)$ (Figure \ref{fig:diap4} (c)).
$G_1$ contains the edges  $(D1,D2)$, whereas the nodes   $(D1-D1)$ have a distance equal to 2 in $G_2$. We identify this condition  as \textbf{gap} that we call \textbf{homogeneous gap}.

Once the edges of the alignment graph are added, a weight is assigned to each edge by applying an ad-hoc scoring function $F$ and the gap threshold $\Delta$. The function assigns a highest score to the matches, respect to mismatch and gaps. The kind of scoring function has a large significance on the resulting alignment graph and on the alignment itself. The algorithm enables the user to choose several values of $\Delta$ to optimize the quality of results. In this work we set the weight assigned to each edge as follows:
homogeneous match equal to $1$, homogeneous mismatch equal to $0.5$, homogeneous gap equal to $0.2$.

\subsection{Topological quality}
Table \ref{tab:ecintra}, Table \ref{tab:ecinter}, Table \ref{tab:ncintra}, Table \ref{tab:ncinter} report the results for synthetic multilayer networks, whereas Table \ref{tab:Recintra}, Table \ref{tab:Recinter}, Table \ref{tab:Rfncintra}, Table \ref{tab:Rfncinter} report the results for real multilayer network.
\begin{table}[!ht]
\centering
\scriptsize
\caption{NCV-G$S^3$ values computed on \textit{intra}-layer edges for all the local alignments built on synthetic networks by applying \textit{Louvain}, \textit{Infomap}, \textit{Greedy} community detection algorithms.}
\scalebox{0.6}{
 \begin{tabular}{|c|c|c|c|c|}
  \hline
  \textbf{Network } & \textbf{Noise} & \textbf{NCV-G$S^3$ with \textit{Louvain} } & \textbf{ NCV-G$S^3$ with \textit{Infomap} }& \textbf{NCV-G$S^3$ with \textit{Greedy} } \\ \hline
  \multirow{6}{*}{N1} &	0	&	0.631	&	0.574	&	0.604	\\	
  &	5	&	0.631	&	0.574	&	0.604	\\	
  &	10	&	0.625	&	0.562	&	0.589	\\	
  &	15	&	0.596	&	0.56	&	0.557	\\	
  &	20	&	0.593	&	0.559	&	0.55	\\	
  &	25	&	0.583	&	0.545	&	0.537	\\	
  \hline
  \multirow{6}{*}{N2} &	0	&	0.635	&	0.599	&	0.6	\\	
  &	5	&	0.635	&	0.599	&	0.6	\\	
  &	10	&	0.628	&	0.581	&	0.583	\\	
  &	15	&	0.592	&	0.54	&	0.58	\\	
  &	20	&	0.573	&	0.527	&	0.576	\\	
  &	25	&	0.559	&	0.527	&	0.533	\\	
  \hline
  \multirow{6}{*}{N3} &	0	&	0.648	&	0.587	&	0.595	\\	
  &	5	&	0.648	&	0.587	&	0.595	\\	
  &	10	&	0.643	&	0.572	&	0.557	\\	
  &	15	&	0.629	&	0.544	&	0.53	\\	
  &	20	&	0.553	&	0.532	&	0.524	\\	
  &	25	&	0.552	&	0.53	&	0.518	\\	
  \hline
  \multirow{6}{*}{N4}
  &	0	&	0.641	&	0.598	&	0.594	\\	
  &	5	&	0.641	&	0.598	&	0.594	\\	
  &	10	&	0.621	&	0.592	&	0.583	\\	
  &	15	&	0.61	&	0.554	&	0.566	\\	
  &	20	&	0.609	&	0.549	&	0.545	\\	
  &	25	&	0.57	&	0.502	&	0.52	\\	
  \hline
  \multirow{6}{*}{N5} 
  & 0	&	0.643	&	0.588	&	0.571	\\	
  &	5	&	0.643	&	0.588	&	0.571	\\	
  &	10	&	0.624	&	0.565	&	0.554	\\	
  &	15	&	0.569	&	0.559	&	0.545	\\	
  &	20	&	0.568	&	0.559	&	0.511	\\	
  &	25	&	0.552	&	0.549	&	0.51	\\	
  \hline
  \multirow{6}{*}{N6}
  &	0	&	0.632	&	0.559	&	0.589	\\	
  &	5	&	0.632	&	0.559	&	0.589	\\	
  &	10	&	0.613	&	0.54	&	0.577	\\	
  &	15	&	0.587	&	0.539	&	0.565	\\	
  &	20	&	0.581	&	0.499	&	0.564	\\	
  &	25	&	0.57	&	0.495	&	0.539	\\	
  \hline
  \multirow{6}{*}{N7}
  &	0	&	0.639	&	0.6	&	0.604	\\	
  &	5	&	0.639	&	0.6	&	0.604	\\	
  &	10	&	0.635	&	0.581	&	0.601	\\	
  &	15	&	0.607	&	0.576	&	0.596	\\	
  &	20	&	0.563	&	0.52	&	0.553	\\	
  &	25	&	0.552	&	0.517	&	0.527	\\	
  \hline
  \multirow{6}{*}{N8}
  &	0	&	0.625	&	0.578	&	0.587	\\	
  &	5	&	0.625	&	0.578	&	0.587	\\	
  &	10	&	0.616	&	0.569	&	0.545	\\	
  &	15	&	0.594	&	0.556	&	0.538	\\	
  &	20	&	0.559	&	0.548	&	0.524	\\	
  &	25	&	0.558	&	0.491	&	0.513	\\	
  \hline
  \multirow{6}{*}{N9}
  &	0	&	0.638	&	0.561	&	0.566	\\	
  &	5	&	0.638	&	0.561	&	0.566	\\	
  &	10	&	0.636	&	0.552	&	0.556	\\	
  &	15	&	0.635	&	0.532	&	0.549	\\	
  &	20	&	0.604	&	0.527	&	0.548	\\	
  &	25	&	0.572	&	0.51	&	0.547	\\	
  \hline
  \multirow{6}{*}{N10}
  &	0	&	0.65	&	0.592	&	0.6 \\
  &	5	&	0.65	&	0.592	&	0.6	\\	
  &	10	&	0.649	&	0.56	&	0.56 \\	
  &	15	&	0.637	&	0.56	&	0.559 \\	
  &	20	&	0.604	&	0.558	&	0.543 \\	
  &	25	&	0.566	&	0.515	&	0.53 \\	
  \hline
  \hline
 \end{tabular}
}
\label{tab:ecintra}
\end{table}

\begin{table}[!ht]
\centering
\scriptsize
\caption{NCV-G$S^3$ values computed on \textit{inter}-layer edges for all the local alignments built on synthetic networks by applying \textit{Louvain}, \textit{Infomap}, \textit{Greedy} community detection algorithms. }
\label{tab:ecinter}
\scalebox{0.6}{
\begin{tabular}{|p{2cm}|p{2cm}|p{2cm}|p{2cm}|p{2cm}|}
\hline

\textbf{Network } & \textbf{Noise} & \textbf{NCV-G$S^3$ with \textit{Louvain} } & \textbf{ NCV-G$S^3$ with \textit{Infomap} }& \textbf{NCV-G$S^3$ with \textit{Greedy} } \\ \hline
\multirow{6}{*}{N1} &	0	&	0.598	&	0.547	&	0.545	\\	
&	5	&	0.598	&	0.547	&	0.545	\\	
&	10	&	0.586	&	0.537	&	0.506	\\	
&	15	&	0.527	&	0.52	&	0.5	\\	
&	20	&	0.504	&	0.487	&	0.5	\\	
&	25	&	0.501	&	0.475	&	0.486	\\	\hline
\multirow{6}{*}{N2} &	0	&	0.583	&	0.524	&	0.546	\\	
&	5	&	0.583	&	0.524	&	0.546	\\	
&	10	&	0.563	&	0.504	&	0.523	\\	
&	15	&	0.555	&	0.502	&	0.516	\\	
&	20	&	0.542	&	0.484	&	0.508	\\	
&	25	&	0.536	&	0.472	&	0.492	\\	\hline
\multirow{6}{*}{N3} &	0	&	0.589	&	0.544	&	0.577	\\	
&	5	&	0.589	&	0.544	&	0.577	\\	
&	10	&	0.589	&	0.526	&	0.556	\\	
&	15	&	0.567	&	0.525	&	0.545	\\	
&	20	&	0.52	&	0.503	&	0.539	\\	
&	25	&	0.516	&	0.48	&	0.527	\\	\hline
\multirow{6}{*}{N4} &	0	&	0.593	&	0.55	&	0.541	\\	
&	5	&	0.593	&	0.55	&	0.541	\\	
&	10	&	0.592	&	0.539	&	0.495	\\	
&	15	&	0.581	&	0.532	&	0.494	\\	
&	20	&	0.507	&	0.515	&	0.482	\\	
&	25	&	0.504	&	0.477	&	0.482	\\	\hline
\multirow{6}{*}{N5} &	0	&	0.598	&	0.532	&	0.567	\\	
&	5	&	0.598	&	0.532	&	0.567	\\	
&	10	&	0.571	&	0.508	&	0.548	\\	
&	15	&	0.546	&	0.5	&	0.498	\\	
&	20	&	0.532	&	0.493	&	0.488	\\	
&	25	&	0.525	&	0.484	&	0.483	\\	\hline
\multirow{6}{*}{N6} &	0	&	0.583	&	0.539	&	0.574	\\	
&	5	&	0.583	&	0.539	&	0.574	\\	
&	10	&	0.564	&	0.509	&	0.549	\\	
&	15	&	0.552	&	0.475	&	0.516	\\	
&	20	&	0.544	&	0.475	&	0.516	\\	
&	25	&	0.525	&	0.464	&	0.489	\\	\hline
\multirow{6}{*}{N7} &	0	&	0.559	&	0.551	&	0.574	\\	
&	5	&	0.559	&	0.551	&	0.574	\\	
&	10	&	0.517	&	0.489	&	0.522	\\	
&	15	&	0.516	&	0.481	&	0.512	\\	
&	20	&	0.514	&	0.474	&	0.496	\\	
&	25	&	0.513	&	0.462	&	0.491	\\	\hline
\multirow{6}{*}{N8} &	0	&	0.586	&	0.557	&	0.577	\\	
&	5	&	0.586	&	0.557	&	0.577	\\	
&	10	&	0.571	&	0.549	&	0.576	\\	
&	15	&	0.537	&	0.528	&	0.556	\\	
&	20	&	0.533	&	0.489	&	0.531	\\	
&	25	&	0.519	&	0.468	&	0.512	\\	\hline
\multirow{6}{*}{N9} &	0	&	0.576	&	0.56	&	0.574	\\	
&	5	&	0.576	&	0.56	&	0.574	\\	
&	10	&	0.564	&	0.558	&	0.555	\\	
&	15	&	0.558	&	0.558	&	0.555	\\	
&	20	&	0.533	&	0.534	&	0.545	\\	
&	25	&	0.515	&	0.525	&	0.487	\\	\hline
\multirow{6}{*}{N10} &	0	&	0.576	&	0.533	&	0.532	\\	
&	5	&	0.576	&	0.533	&	0.532	\\	
&	10	&	0.524	&	0.53	&	0.519	\\	
&	15	&	0.521	&	0.5	&	0.519	\\	
&	20	&	0.515	&	0.499	&	0.493	\\	
&	25	&	0.504	&	0.468	&	0.49	\\	\hline
\hline
\end{tabular}}
\end{table}

\begin{table}[!ht]
\centering
\scriptsize
\caption{F-NC values computed on \textit{intra}-layer edges for all the local alignments built on synthetic networks by applying \textit{Louvain}, \textit{Infomap}, \textit{Greedy} community detection algorithms. }
\label{tab:ncintra}
\scalebox{0.6}{
\begin{tabular}{|p{2cm}|p{2cm}|p{2cm}|p{2cm}|p{2cm}|}
\hline

\textbf{Network } & \textbf{Noise} & \textbf{F-NC with \textit{Louvain} } & \textbf{ F-NC with \textit{Infomap} }& \textbf{F-NC with \textit{Greedy} } \\ \hline

\multirow{6}{*}{N1} &	0	&	0.999	&	0.934	&	0.974	\\	
&	5	&	0.999	&	0.934	&	0.974	\\	
&	10	&	0.994	&	0.913	&	0.927	\\	
&	15	&	0.961	&	0.912	&	0.889	\\	
&	20	&	0.926	&	0.86	&	0.87	\\	
&	25	&	0.896	&	0.857	&	0.863	\\	\hline
\multirow{6}{*}{N2} &	0	&	0.986	&	0.863	&	0.971	\\	
&	5	&	0.986	&	0.863	&	0.971	\\	
&	10	&	0.957	&	0.768	&	0.907	\\	
&	15	&	0.951	&	0.729	&	0.857	\\	
&	20	&	0.927	&	0.63	&	0.76	\\	
&	25	&	0.896	&	0.618	&	0.71	\\	\hline
\multirow{6}{*}{N3} &	0	&	0.981	&	0.87	&	0.93	\\	
&	5	&	0.981	&	0.87	&	0.93	\\	
&	10	&	0.96	&	0.836	&	0.859	\\	
&	15	&	0.919	&	0.821	&	0.858	\\	
&	20	&	0.902	&	0.741	&	0.804	\\	
&	25	&	0.883	&	0.613	&	0.799	\\	\hline
\multirow{6}{*}{N4} &	0	&	0.997	&	0.891	&	0.921	\\	
&	5	&	0.997	&	0.891	&	0.921	\\	
&	10	&	0.992	&	0.889	&	0.879	\\	
&	15	&	0.945	&	0.835	&	0.825	\\	
&	20	&	0.927	&	0.791	&	0.785	\\	
&	25	&	0.893	&	0.683	&	0.747	\\	\hline
\multirow{6}{*}{N5} &	0	&	0.963	&	0.951	&	0.92	\\	
&	5	&	0.963	&	0.951	&	0.92	\\	
&	10	&	0.907	&	0.853	&	0.777	\\	
&	15	&	0.905	&	0.852	&	0.768	\\	
&	20	&	0.895	&	0.82	&	0.753	\\	
&	25	&	0.894	&	0.684	&	0.733	\\	\hline
\multirow{6}{*}{N6} &	0	&	0.938	&	0.917	&	0.877	\\	
&	5	&	0.938	&	0.917	&	0.877	\\	
&	10	&	0.928	&	0.905	&	0.868	\\	
&	15	&	0.914	&	0.816	&	0.86	\\	
&	20	&	0.891	&	0.699	&	0.854	\\	
&	25	&	0.881	&	0.619	&	0.724	\\	\hline
\multirow{6}{*}{N7} &	0	&	0.998	&	0.967	&	0.908	\\	
&	5	&	0.998	&	0.967	&	0.908	\\	
&	10	&	0.985	&	0.922	&	0.906	\\	
&	15	&	0.977	&	0.828	&	0.861	\\	
&	20	&	0.938	&	0.765	&	0.846	\\	
&	25	&	0.892	&	0.764	&	0.723	\\	\hline
\multirow{6}{*}{N8} &	0	&	0.96	&	0.97	&	0.932	\\	
&	5	&	0.96	&	0.97	&	0.932	\\	
&	10	&	0.945	&	0.932	&	0.881	\\	
&	15	&	0.928	&	0.797	&	0.864	\\	
&	20	&	0.893	&	0.755	&	0.755	\\	
&	25	&	0.884	&	0.709	&	0.721	\\	\hline
\multirow{6}{*}{N9} &	0	&	0.992	&	0.916	&	0.961	\\	
&	5	&	0.992	&	0.916	&	0.961	\\	
&	10	&	0.989	&	0.786	&	0.931	\\	
&	15	&	0.974	&	0.739	&	0.925	\\	
&	20	&	0.962	&	0.736	&	0.884	\\	
&	25	&	0.915	&	0.651	&	0.836	\\	\hline
\multirow{6}{*}{N10} &	0	&	0.986	&	0.967	&	0.894	\\	
&	5	&	0.986	&	0.967	&	0.894	\\	
&	10	&	0.971	&	0.832	&	0.834	\\	
&	15	&	0.963	&	0.798	&	0.806	\\	
&	20	&	0.94	&	0.792	&	0.766	\\	
&	25	&	0.914	&	0.704	&	0.736	\\	\hline
\hline
\end{tabular}}
\end{table}

\begin{table}[!ht]
\centering
\scriptsize
\caption{F-NC values computed on \textit{inter}-layer edges for all the local alignments built on synthetic networks by applying \textit{Louvain}, \textit{Infomap}, \textit{Greedy} community detection algorithms. }
\label{tab:ncinter}
\scalebox{0.6}{
\begin{tabular}{|p{2cm}|p{2cm}|p{2cm}|p{2cm}|p{2cm}|}
\hline

\textbf{Network } & \textbf{Noise} & \textbf{F-NC with \textit{Louvain} } & \textbf{ F-NC with \textit{Infomap} }& \textbf{F-NC with \textit{Greedy} } \\ \hline
\multirow{6}{*}{N1} &	0	&	0.771	&	0.706	&	0.757	\\	
&	5	&	0.771	&	0.706	&	0.757	\\	
&	10	&	0.77	&	0.671	&	0.7	\\	
&	15	&	0.717	&	0.64	&	0.678	\\	
&	20	&	0.659	&	0.614	&	0.652	\\	
&	25	&	0.645	&	0.604	&	0.592	\\	\hline
\multirow{6}{*}{N2} &	0	&	0.796	&	0.698	&	0.743	\\	
&	5	&	0.796	&	0.698	&	0.743	\\	
&	10	&	0.723	&	0.695	&	0.7	\\	
&	15	&	0.69	&	0.68	&	0.691	\\	
&	20	&	0.674	&	0.605	&	0.681	\\	
&	25	&	0.607	&	0.569	&	0.623	\\	\hline
\multirow{6}{*}{N3} &	0	&	0.793	&	0.696	&	0.728	\\	
&	5	&	0.793	&	0.696	&	0.728	\\	
&	10	&	0.697	&	0.692	&	0.704	\\	
&	15	&	0.677	&	0.661	&	0.699	\\	
&	20	&	0.668	&	0.65	&	0.69	\\	
&	25	&	0.614	&	0.566	&	0.585	\\	\hline
\multirow{6}{*}{N4} &	0	&	0.763	&	0.666	&	0.775	\\	
&	5	&	0.763	&	0.666	&	0.775	\\	
&	10	&	0.724	&	0.621	&	0.74	\\	
&	15	&	0.679	&	0.621	&	0.698	\\	
&	20	&	0.639	&	0.588	&	0.653	\\	
&	25	&	0.6	&	0.586	&	0.601	\\	\hline
\multirow{6}{*}{N5} &	0	&	0.745	&	0.732	&	0.775	\\	
&	5	&	0.745	&	0.732	&	0.775	\\	
&	10	&	0.702	&	0.697	&	0.683	\\	
&	15	&	0.666	&	0.694	&	0.607	\\	
&	20	&	0.658	&	0.628	&	0.603	\\	
&	25	&	0.616	&	0.572	&	0.585	\\	\hline
\multirow{6}{*}{N6} &	0	&	0.748	&	0.678	&	0.745	\\	
&	5	&	0.748	&	0.678	&	0.745	\\	
&	10	&	0.738	&	0.678	&	0.727	\\	
&	15	&	0.697	&	0.642	&	0.625	\\	
&	20	&	0.62	&	0.627	&	0.603	\\	
&	25	&	0.612	&	0.591	&	0.593	\\	\hline
\multirow{6}{*}{N7} &	0	&	0.735	&	0.718	&	0.777	\\	
&	5	&	0.735	&	0.718	&	0.777	\\	
&	10	&	0.728	&	0.67	&	0.713	\\	
&	15	&	0.692	&	0.659	&	0.698	\\	
&	20	&	0.661	&	0.592	&	0.631	\\	
&	25	&	0.608	&	0.561	&	0.602	\\	\hline
\multirow{6}{*}{N8} &	0	&	0.758	&	0.719	&	0.753	\\	
&	5	&	0.758	&	0.719	&	0.753	\\	
&	10	&	0.739	&	0.7	&	0.743	\\	
&	15	&	0.718	&	0.695	&	0.735	\\	
&	20	&	0.707	&	0.669	&	0.699	\\	
&	25	&	0.627	&	0.597	&	0.683	\\	\hline
\multirow{6}{*}{N9} &	0	&	0.712	&	0.708	&	0.763	\\	
&	5	&	0.712	&	0.708	&	0.763	\\	
&	10	&	0.665	&	0.687	&	0.735	\\	
&	15	&	0.659	&	0.685	&	0.724	\\	
&	20	&	0.658	&	0.669	&	0.714	\\	
&	25	&	0.621	&	0.619	&	0.666	\\	\hline
\multirow{6}{*}{N10} &	0	&	0.765	&	0.724	&	0.764	\\	
&	5	&	0.765	&	0.724	&	0.764	\\	
&	10	&	0.754	&	0.628	&	0.761	\\	
&	15	&	0.692	&	0.612	&	0.725	\\	
&	20	&	0.681	&	0.571	&	0.712	\\	
&	25	&	0.618	&	0.542	&	0.691	\\	\hline
\hline
\end{tabular}}
\end{table}

\begin{table}[!ht]
\centering
\scriptsize
\caption{NCV-G$S^3$ values computed on \textit{intra}-layer edges for all the local alignments built on real network by applying \textit{Louvain}, \textit{Infomap}, \textit{Greedy} community detection algorithms. }
\label{tab:Recintra}
\scalebox{0.6}{
\begin{tabular}{|p{2cm}|p{2cm}|p{2cm}|p{2cm}|p{2cm}|}
\hline

\textbf{Network } & \textbf{Noise} & \textbf{NCV-G$S^3$ with \textit{Louvain} } & \textbf{ NCV-G$S^3$ with \textit{Infomap} }& \textbf{NCV-G$S^3$ with \textit{Greedy} } \\ \hline
\multirow{6}{*}{N1} &	0	&	0.884	&	0.565	&	0.773	\\	
&	5	&	0.884	&	0.565	&	0.773	\\	
&	10	&	0.879	&	0.554	&	0.769	\\	
&	15	&	0.877	&	0.551	&	0.763	\\	
&	20	&	0.836	&	0.549	&	0.713	\\	
&	25	&	0.818	&	0.541	&	0.702	\\	\hline

\hline
\end{tabular}}
\end{table}

\begin{table}[!ht]
\centering
\scriptsize
\caption{NCV-G$S^3$ values computed on \textit{inter}-layer edges for all the local alignments built on real network by applying \textit{Louvain}, \textit{Infomap}, \textit{Greedy} community detection algorithms. }
\label{tab:Recinter}
\scalebox{0.6}{
\begin{tabular}{|p{2cm}|p{2cm}|p{2cm}|p{2cm}|p{2cm}|}
\hline

\textbf{Network } & \textbf{Noise} & \textbf{NCV-G$S^3$ with \textit{Louvain} } & \textbf{ NCV-G$S^3$ with \textit{Infomap} }& \textbf{NCV-G$S^3$ with \textit{Greedy} } \\ \hline
\multirow{6}{*}{Real Network} &	0	&	0.661	&	0.565	&	0.635	\\	
&	5	&	0.661	&	0.565	&	0.635	\\	
&	10	&	0.65	&	0.565	&	0.634	\\	
&	15	&	0.622	&	0.558	&	0.609	\\	
&	20	&	0.602	&	0.555	&	0.609	\\	
&	25	&	0.592	&	0.544	&	0.599	\\	\hline

\hline
\end{tabular}}
\end{table}

\begin{table}[!ht]
\centering
\scriptsize
\caption{F-NC values computed on \textit{intra}-layer edges forfor all the local alignments built on real network by applying \textit{Louvain}, \textit{Infomap}, \textit{Greedy} community detection algorithms. }
\label{tab:Rfncintra}
\scalebox{0.6}{
\begin{tabular}{|p{2cm}|p{2cm}|p{2cm}|p{2cm}|p{2cm}|}
\hline

\textbf{Network } & \textbf{Noise} & \textbf{F-NC with \textit{Louvain} } & \textbf{ F-NC with \textit{Infomap} }& \textbf{F-NC with \textit{Greedy} } \\ \hline

\multirow{6}{*}{Real Network} &	0	&	0.661	&	0.565	&	0.635	\\	
&	5	&	0.661	&	0.565	&	0.635	\\	
&	10	&	0.65	&	0.565	&	0.634	\\	
&	15	&	0.622	&	0.558	&	0.609	\\	
&	20	&	0.602	&	0.555	&	0.609	\\	
&	25	&	0.592	&	0.544	&	0.599	\\	\hline

\hline
\end{tabular}}
\end{table}

\begin{table}[!ht]
\centering
\scriptsize
\caption{F-NC values computed on \textit{inter}-layer edges for all the local alignments built on real network by applying \textit{Louvain}, \textit{Infomap}, \textit{Greedy} community detection algorithms. }
\label{tab:Rfncinter}
\scalebox{0.6}{
\begin{tabular}{|p{2cm}|p{2cm}|p{2cm}|p{2cm}|p{2cm}|}
\hline

\textbf{Network } & \textbf{Noise} & \textbf{F-NC with \textit{Louvain} } & \textbf{ F-NC with \textit{Infomap} }& \textbf{F-NC with \textit{Greedy} } \\ \hline
\multirow{6}{*}{Real Network} &	0	&	0.576	&	0.497	&	0.528	\\	
&	5	&	0.576	&	0.497	&	0.528	\\	
&	10	&	0.538	&	0.486	&	0.521	\\	
&	15	&	0.511	&	0.484	&	0.51	\\	
&	20	&	0.507	&	0.478	&	0.507	\\	
&	25	&	0.498	&	0.46	&	0.486	\\	\hline

\hline
\end{tabular}}
\end{table}
    
\end{document}